\documentstyle[aps,epsfig]{revtex}
\begin{document}
\draft
\title{Calculation of the chiral Lagrangian coefficients from the underlying 
\\theory of QCD: A simple approach}
\vspace{0.2cm}
\author{Hua Yang$^{b,d}$,~~~~ Qing Wang$^{a,b,c}$,~~~~ Yu-Ping Kuang$^{a,b,c}$,
~~~~Qin Lu$^b$}
\vspace{0.2cm}
\address{
a. CCAST (World Laboratory),
P.O.Box 8730, Beijing 100080, China \\
b. Department of Physics, Tsinghua University, Beijing 100084, China
\footnote{Mailing address}\\
c. Institute of Theoretical Physics, Academia Sinica, Beijing 100080, China\\
d. Institute of Electronic Technology, Information Engineering
University, Zhengzhou 450004, Henan, China}
\date{TUHEP-TH-02132}

\maketitle

\begin{abstract}
We calculate the coefficients in the chiral Lagrangian approximately
from QCD based on a previous study of deriving the chiral Lagrangian
from the first principles of QCD in which the chiral Lagrangian
coefficients are defined in terms of certain Green's functions in QCD.
We first show that, in the large-$N_c$ limit, the anomaly part contributions 
to the coefficients are exactly cancelled by certain terms in the normal part
contributions, and the final results of the coefficients only
concern the remaining normal part contributions depending on QCD interactions.
We then do the calculation in a simple approach
with the approximations of taking the large-$N_c$ limit, the leading order in 
dynamical perturbation theory, and the improved ladder approximation, thereby 
the relevant Green's functions are expressed in terms of the quark self energy 
$\Sigma(p^2)$. By solving the Schwinger-Dyson equation for
$\Sigma(p^2)$, we obtain the approximate QCD predicted coefficients
and quark condensate which are consistent with the experimental values.
\end{abstract}

\bigskip
PACS number(s): 12.39.Fe, 11.30.Rd, 12.38.Aw, 12.38.Lg

\vspace{1cm} 
\section{INTRODUCTION}

Because of its nonperturbative nature, studying low energy hadron physics
in QCD is a long standing difficult problem. For low lying pseudoscalar 
mesons, a widely used approach is the theory of the effective chiral Lagrangian 
based on the consideration of the global symmetry of the system and the 
momentum expansion without dealing with the nonperturbative dynamics of QCD 
\cite{weinberg}\cite{GS}. In the chiral Lagrangian approach, the
coefficients in the Lagrangian are all unknown phenomenological 
parameters which should be determined by experimental inputs. 
The number of unknown parameters increases rapidly with the increase of 
the precision in the momentum expansion. 
Therefore studying the relation between the chiral Langrangian 
and the fundamental principles of QCD will not only be theoretically 
interesting for a deeper understanding of the chiral Lagrangian, but will also
be helpful for reducing the number of unknown parameters and increasing the 
predictive power of the chiral Lagrangian.

In a previous paper, Ref.\cite{WKWX1}, certain techniques were developed,
with which the chiral Lagrangian was formally derived from the  
first principles of QCD without taking approximations. The chiral
Lagrangian coefficients are contributed both by the anomaly part
(from the quark functional measure) and the normal part
(from the QCD Lagrangian). In Ref.\cite{WKWX1},
all the chiral Lagrangian coefficients contributed from the normal part of 
the theory are expressed in terms of certain Green's
functions in QCD, which can be regarded as exact QCD definitions of the chiral
Lagrangian coefficients. 
After expanding the effective action in powers of the
rotated sources (momentum expansion), the effective action , up to $O(p^4)$, 
contributed from the normal part is of the form \cite{WKWX1}
\begin{eqnarray}                             
S_{\rm eff}^{{(\rm norm})}
&=&\int d^4x~{\rm  tr}_f\bigg[F_0^2a_{\Omega}^2+F_0^2B_0s_{\Omega}
-{\cal K}_1[d_{\mu}a_{\Omega}^{\mu}]^2 
-{\cal K}_2(d^{\mu}a_{\Omega}^{\nu}-d^{\nu}a_{\Omega}^{\mu})
(d_{\mu}a_{\Omega,\nu}-d_{\nu}a_{\Omega,\mu})
\nonumber\\
&&\hspace{0.5cm}
+{\cal K}_3[a_{\Omega}^2]^2
+{\cal K}_4a_{\Omega}^{\mu}a_{\Omega}^{\nu}a_{\Omega,\mu}a_{\Omega,\nu}
+{\cal K}_5a_\Omega^2{\rm tr}_f[a_{\Omega}^2]
+{\cal K}_6a_{\Omega}^{\mu}a_{\Omega}^{\nu}
{\rm tr}_f[a_{\Omega,\mu}a_{\Omega,\nu}]+{\cal K}_7s_{\Omega}^2
\nonumber\\
&&\hspace{0.5cm}
+{\cal K}_8s_{\Omega}{\rm tr}_f[s_{\Omega}]
+{\cal K}_9p_{\Omega}^2
+{\cal K}_{10}p_{\Omega}{\rm tr}_f[p_{\Omega}]
+{\cal K}_{11}s_{\Omega}a_{\Omega}^2+{\cal K}_{12}s_{\Omega}{\rm tr}_f
[a_{\Omega}^2]
-{\cal K}_{13}V_{\Omega}^{\mu\nu}V_{\Omega,\mu\nu}
\nonumber\\
&&\hspace{0.5cm}
+i{\cal K}_{14}V_{\Omega}^{\mu\nu}a_{\Omega,\mu}a_{\Omega,\nu}
+{\cal K}_{15}p_{\Omega}d_{\mu}a^{\mu}_{\Omega}\bigg] +O(p^6),
\label{p4}
\end{eqnarray}
where $\Omega$ is related to the nonlinearly realized meson field $U$ by 
$U=\Omega^2$; $s_\Omega,
~p_\Omega,~v_\Omega$, and $a_\Omega$ are, respectively, the external scalar, 
pseudoscalar, vector, and axial-vector sources rotated by $\Omega$; and the 
${\cal K}$s are terms with different Lorentz structures in the relevant QCD 
Green's functions.
The obtained expressions for the chiral Lagrangian 
coefficients are

\vspace{0.2cm}
\null\noindent
$O(p^2)$:
\begin{eqnarray}                              
&&F^2_0=\frac{i}{8(N_f^2-1)}\int~d^4x~\bigg[\langle [\bar{\psi}^a(0)\gamma^\mu
\gamma_5\psi^b(0)][\bar{\psi}^b(x)\gamma_\mu\gamma_5\psi^a(x)]\rangle
-\frac{1}{N_f}\langle [\bar{\psi}^a(0)\gamma^\mu\gamma_5
\psi^a(0)]
[\bar{\psi}^b(x)\gamma_\mu\gamma_5\psi^b(x)]\rangle\nonumber\\
&&\hspace{0.8cm}-\langle \bar{\psi}^a(0)\gamma^\mu\gamma_5\psi^b(0)\rangle
\langle\bar{\psi}^b(x)\gamma_\mu\gamma_5\psi^a(x)\rangle
+\frac{1}{N_f}\langle \bar{\psi}^a(0)\gamma^\mu\gamma_5\psi^a(0)\rangle
\langle\bar{\psi}^b(x)\gamma_\mu\gamma_5\psi^b(x)\rangle\bigg]
,\nonumber\\
&&F_0^2B_0=-\frac{1}{N_f}\langle \bar{\psi}\psi\rangle,
\label{F_0B_0}
\end{eqnarray}

\null\noindent
$O(p^4)$:
\begin{eqnarray}                          
&&L_1^{(\rm norm)}
=\frac{1}{32}{\cal K}_4+\frac{1}{16}{\cal K}_5+\frac{1}{16}{\cal K}_{13}
-\frac{1}{32}{\cal K}_{14},
~~~~~~~~~~~L_2^{(\rm norm)}
=\frac{1}{16}({\cal K}_4+{\cal K}_6)+\frac{1}{8}{\cal K}_{13}-\frac{1}{16}
{\cal K}_{14},
\nonumber\\
&&L_3^{(\rm norm)}
=\frac{1}{16}({\cal K}_3-2{\cal K}_4-6{\cal K}_{13}+3{\cal K}_{14}),
~~~~~~~~~~~~~~~L_4^{(\rm norm)}
=\frac{{\cal K}_{12}}{16B_0},
\nonumber\\
&&L_5^{(\rm norm)}
=\frac{{\cal K}_{11}}{16B_0},
~~~~~~~~~~~~~~~~~~~~~~~~~~~~~~~~~~~~~~~~~~~~~~L_6^{(\rm norm)}
=\frac{{\cal K}_8}{16B_0^2},
\nonumber\\
&&L_7^{(\rm norm)}
=-\frac{{\cal K}_1}{16N_f}-\frac{{\cal K}_{10}}{16B_0^2}
-\frac{{\cal K}_{15}}{16B_0N_f},
~~~~~~~~~~~~~~~~~L_8^{(\rm norm)}
=\frac{1}{16}[{\cal K}_1+\frac{1}{B_0^2}{\cal K}_7
-\frac{1}{B_0^2}{\cal K}_9
+\frac{1}{B_0}{\cal K}_{15}],
\nonumber\\
&&L_9^{(\rm norm)}
= \frac{1}{8}(4{\cal K}_{13}-{\cal K}_{14}),
~~~~~~~~~~~~~~~~~~~~~~~~~~~~~~~~~~L_{10}^{(\rm norm)}
=\frac{1}{2}({\cal K}_2-{\cal K}_{13}),
\nonumber\\
&&H_1^{(\rm norm)}
=-\frac{1}{4}({\cal K}_2+{\cal K}_{13}),
~~~~~~~~~~~~~~~~~~~~~~~~~~~~~~~~~~H_2^{(\rm norm)}
=\frac{1}{8}[-{\cal K}_1+\frac{1}{B_0^2}{\cal K}_7
+\frac{1}{B_0^2}{\cal K}_9-\frac{1}{B_0}{\cal K}_{15}].\label{p4C}
\end{eqnarray}
Together with the anomaly part contributions, the complete coefficients 
are given by
\begin{eqnarray}                       
L_i=L_i^{(\rm anom)}+L_i^{(\rm norm)},~~~~i=1,\cdots,10,
~~~~~~~~~~~~~~~~H_i=H_i^{(\rm anom)}+H_i^{(\rm norm)},~~~~i=1,2,
\label{totalL}
\end{eqnarray}
where the superscripts (anom) and (norm) denote the anomaly part and 
normal part contributions, respectively.
 
In the literature, the anomaly part contributions are usually
calculated by means of the heat kernel regularization technique \cite{Espriu}.
However, this technique is difficult to implement in the calculation of
the normal part contributions which contain complicated functions of the 
momentum, say the quark self-energy $\Sigma(p^2)$, reflecting nonperturbative 
QCD dynamics (which are even unspecified in the analytical part of the 
calculation). In order to treat the anomaly part and the normal part 
contributions on equal footing, a certain new regularization technique
feasible for the calculations of both parts should be developed.
In this paper, we use the generalized Schwinger proper time
regularization technique developed in Ref.\cite{det} to regularize the system, 
which keeps the local chiral symmetry at every step in the calculation, and 
can be applied to the calculations of both the contributions from the anomaly 
part and from the normal part. Thus, in this paper, the contributions from the 
anomaly part and the normal part are calculated by means of the same 
technique. As the first conclusion of this unified treatment, we show that 
{\it the anomaly contributions to the chiral Lagrangian coefficients given in 
Ref.\cite{Espriu}, which are independent of QCD interactions, will actually be 
cancelled by certain terms in the normal part contributions, and the final 
expressions for the coefficients concern only the remaining terms from the 
normal part contributions related to QCD interactions}. It should be so since 
the coefficients indicate meson interactions which should be residual 
interactions between quarks and gluons, and thus should depend on QCD 
interactions. These contributions have not been carefully calculated in the 
literature. It has been shown in Ref.\cite{WKWX1} that in the approximations 
of large-$N_c$ limit, leading order in dynamical perturbation, and
improved ladder approximation, the formula for $F_0^2$ in Eqs.(\ref{F_0B_0})
reduces to the well-known Pagels-Stokar formula \cite{PS}
in which all dynamical effects from QCD are represented by the quark 
self-energy $\Sigma(p^2)$ in the formula. In this paper, we take the 
same approximations to calculate the chiral Lagrangian coefficients (the 
relevant QCD Green's functions) as an illustration of the main feature of how
QCD predicts the chiral Lagrangian coefficients. Similar to the case of the 
Pagels-Stokar formula, the relevant QCD Green's functions can all be 
expressed as functions of the quark self-energy $\Sigma(p^2)$. By solving the 
Schwinger-Dyson equation, we obtain $\Sigma(p^2)$, and thus the approximate 
QCD predicted values of the coefficients. We shall see that the obtained coefficients 
$L_1,\cdots,L_{10}$ and quark condensate are consistent with the experimental 
values.
The calculation is checked by the absence of divergences in the large-$N_c$ 
limit as it should be since the divergent meson-loop contributions are of 
next-to-the-leading order in the $1/N_c$ expansion. 
 Although the present approximation is rather
crude, {\it it reveals the main feature of QCD predictions for the chiral
Lagrangian coefficients}.

This paper is organized as follows: In Sec. II, we calculate the anomaly part 
contributions to the $O(p^4)$ coefficients using the Schwinger proper
time regularization technique, and the results coincide with those in 
Ref.\cite{Espriu} in the chiral limit.
Then, in Sec. III, we apply the same technique to the normal part, and
show generally that, in the large-$N_c$ limit, the anomaly part
contributions to the chiral Lagrangian coefficients are exactly cancelled by 
the contributions from a piece in the normal part independent of the 
quark self-energy, and the contributions from the remaining piece in the 
normal part depending on the quark self-energy play the real role in the 
chiral Lagrangian coefficients. 
Specific approximations in the calculation of the normal part
contributions and the formulae for the complete chiral Lagrangian coefficients 
in terms of the quark self-energy are given in Sec. IV. In Sec. V, we present 
the numerical calculations of the quark self-energy and the obtained values of 
the chiral Lagrangian coefficients. Section VI is a concluding remark.

\section{On the Contributions from the Anomaly part}

In order to see the relation between the anomaly part and the normal
part contributions to the chiral Lagrangian coefficients, we present
here the calculation of the anomaly part contributions by means of the 
Schwinger proper time regularization. We shall see that the obtained
results exactly coincide with those obtained from the heat kernel
technique \cite{Espriu}.
Our present approach is different from that in Ref.\cite{Espriu} in the sense 
that the constant constituent quark mass $M_Q$ is not put in by hand as
is done in Ref.\cite{Espriu} but is naturally included in the normal part
solution through the dynamical quark mass reflecting chiral symmetry breaking.
Therefore our result of the anomaly part contribution is to compare with that 
in Ref.\cite{Espriu} in the chiral limit.

In the Schwinger proper time regularization, 
the anomaly part does not contribute to
the coefficients of the $O(p^2)$ terms in the case corresponding to the result with 
$M_Q=0$ in Ref.\cite{Espriu}\footnote{If one takes a momentum cutoff 
$\Lambda$ to regularize the divergent integrals as was done in 
Ref.\cite{Espriu} before putting in the constituent quark mass $M_Q$, the 
$O(p^2)$ coefficient $F_0^2$ will be proportional to $\Lambda^2$ (cf. 
Ref.\cite{Espriu}). As has been pointed out in Ref.\cite{WKWX1}
this term is exactly cancelled by a corresponding term in the normal part 
contribution [cf. Eq.(74) in Ref.\cite{WKWX1}].}. Therefore we are only going 
to calculate the anomaly contribution to the coefficients of the $O(p^4)$ 
terms.

The anomaly  term in the path-integral formalism is
\begin{eqnarray}                                      
S_{\rm eff}^{({\rm anom})}&\equiv&-i\times{\rm anomaly~~ terms}=
-iN_c[{\rm Tr}\ln(i\partial\!\!\!\! /\;+J)
-{\rm Tr}\ln(i\partial\!\!\!\! /\;+J_{\Omega})]\nonumber\\
&=&iN_c[{\rm Tr}\ln(i\partial\!\!\!\! /\;+J_{\Omega})
+\Omega-{\rm independent~term}].
\label{Sanom}
\end{eqnarray}
The $\Omega$-independent term is independent of the $U$ field, so that it is 
irrelevant to the chiral Lagrangian coefficients. We shall only evaluate the 
$\Omega$-dependent term in Eq.(\ref{Sanom}). To have a 
unified parametrization, we can parametrize the anomaly
contributed effective action similar to that in Eq.(\ref{p4}), i.e.,
\begin{eqnarray}                                        
S_{\rm eff}^{({\rm anom})}&=&\int d^4x~{\rm  tr}_f\bigg[
-{\cal K}_1^{\rm (anom)}[d_{\mu}a_{\Omega}^{\mu}]^2 
-{\cal K}_2^{\rm (anom)}(d^{\mu}a_{\Omega}^{\nu}-d^{\nu}a_{\Omega}^{\mu})
(d_{\mu}a_{\Omega,\nu}-d_{\nu}a_{\Omega,\mu})
+{\cal K}_3^{\rm (anom)}[a_{\Omega}^2]^2
\nonumber\\
&&+{\cal K}_4^{\rm (anom)}a_{\Omega}^{\mu}a_{\Omega}^{\nu}
a_{\Omega,\mu}a_{\Omega,\nu}
+{\cal K}_5^{\rm (anom)}a_\Omega^2{\rm tr}_f[a_{\Omega}^2]
+{\cal K}_6^{\rm (anom)}a_{\Omega}^{\mu}a_{\Omega}^{\nu}
{\rm tr}_f[a_{\Omega,\mu}a_{\Omega,\nu}]+{\cal K}_7^{\rm (anom)}s_{\Omega}^2
\nonumber\\
&&+{\cal K}_8^{\rm (anom)}s_{\Omega}{\rm tr}_f[s_{\Omega}]
+{\cal K}_9^{\rm (anom)}p_{\Omega}^2
+{\cal K}_{10}^{\rm (anom)}p_{\Omega}{\rm tr}_f[p_{\Omega}]
+{\cal K}_{11}^{\rm (anom)}s_{\Omega}a_{\Omega}^2
\nonumber\\
&&+{\cal K}_{12}^{\rm (anom)}s_{\Omega}{\rm tr}_f[a_{\Omega}^2]
-{\cal K}_{13}^{\rm (anom)}V_{\Omega}^{\mu\nu}V_{\Omega,\mu\nu}
+i{\cal K}_{14}^{\rm (anom)}V_{\Omega}^{\mu\nu}a_{\Omega,\mu}a_{\Omega,\nu}
+{\cal K}_{15}^{\rm (anom)}p_{\Omega}d_{\mu}a^{\mu}_{\Omega}\bigg] 
\nonumber\\
&&+O(p^6)+U-\mbox{independent~source~terms}.
\label{p4anom}
\end{eqnarray}
The $\Omega$-dependent term in Eq.(\ref{Sanom}) suffers from ultraviolet 
divergence, and we take the Schwinger proper time regularization with
an ultraviolet cutoff parameter $\Lambda$ to regularize it. To apply this 
regularization, we first work in the Euclidean space-time, and analytically 
continue the results to the Minkowskian space-time after the evaluation.
The main procedure of evaluating the general functional determinant
including the quark self-energy $\Sigma$ is described in APPENDIX A. In
the case of $S^{(\rm anom)}_{\rm eff}$, there is no $\Sigma$-dependence
in Eq.(\ref{Sanom}). However, for regularizing the infrared divergence, we
should replace the $\Sigma$ in Eqs.(\ref{lnSigma}) and (\ref{momexp}) by an 
infrared cutoff parameter $\kappa$. The momentum integration in 
Eq.(\ref{momexp}) can be explicitly carried out with a lengthy but elementary 
calculation. After expanding in powers of the external sources, we can 
identify the expressions for ${\cal K}_1^{(\rm anom)},\cdots,
{\cal K}_{15}^{(\rm anom)}$ by comparing with the form of Eq.(\ref{p4anom}), 
and we obtain 
\begin{eqnarray}                                         
&&{\cal K}_1^{\rm (anom)}=-\frac{N_c}{24\pi^2},
~~~~~~~~~~~~~~~~~~~~~~~~~~~~~~~~~~~~~~~~~~~~
{\cal K}_2^{\rm (anom)}=-\frac{N_c}{48\pi^2}\lim_{\kappa
\rightarrow 0}
\lim_{\Lambda\rightarrow\infty}(\ln\frac{\kappa^2}{\Lambda^2}+\gamma+1),
\nonumber\\
&&{\cal K}_3^{\rm (anom)}=\frac{N_c}{24\pi^2}\lim_{\kappa\rightarrow 0}
\lim_{\Lambda\rightarrow\infty}(\ln\frac{\kappa^2}{\Lambda^2}+\gamma+4),
~~~~~~~~~~~~~~{\cal K}_4^{\rm (anom)}=-\frac{N_c}{24\pi^2}\lim_{\kappa\rightarrow 
0}\lim_{\Lambda\rightarrow\infty}(\ln\frac{\kappa^2}{\Lambda^2}+\gamma+2),
\nonumber\\
&&{\cal K}_5^{\rm (anom)}={\cal K}_6^{\rm (anom)}=0,
~~~~~~~~~~~~~~~~~~~~~~~~~~~~~~~~~~~~~~
{\cal K}_7^{\rm (anom)}={\cal K}_9^{\rm (anom)}=\frac{N_c}{8\pi^2}
\lim_{\Lambda\rightarrow\infty}\Lambda^2,
\nonumber\\
&&{\cal K}_8^{\rm (anom)}={\cal K}_{10}^{\rm (anom)}=
{\cal K}_{11}^{\rm (anom)}=
{\cal K}_{12}^{\rm (anom)}=0,
~~~~~~~~~~{\cal K}_{13}^{\rm (anom)}=-\frac{N_c}{48\pi^2}\lim_{\kappa
\rightarrow 0}
\lim_{\Lambda\rightarrow\infty}(\ln\frac{\kappa^2}{\Lambda^2}+\gamma),
\nonumber\\
&&{\cal K}_{14}^{\rm (anom)}=-\frac{N_c}{12\pi^2}\lim_{\kappa\rightarrow 0}
\lim_{\Lambda\rightarrow\infty}(\ln\frac{\kappa^2}{\Lambda^2}+\gamma+2),
~~~~~~~~~~~~~{\cal K}_{15}^{\rm (anom)}=0.
\label{anomK}
\end{eqnarray}

Comparing with the standard form of momentum expansion to identify the 
$O(p^4)$ chiral Lagrangian coefficients, we obtain the anomaly 
contribution to these coefficients
\begin{eqnarray}                          
&&L_1^{\rm (anom)}
=\frac{N_c}{384\pi^2},~~~~~~~~~~
L_2^{\rm (anom)}
=\frac{N_c}{192\pi^2},~~~~~~~~~~
L_3^{\rm (anom)}
=-\frac{N_c}{96\pi^2}
~~~~~~~~~L_4^{\rm (anom)}=L_5^{\rm (anom)}=L_6^{\rm (anom)}=0
\nonumber,\\
&&L_7^{\rm (anom)}
=\frac{N_c}{1152\pi^2},~~~~~~~~L_8^{\rm (anom)}
=-\frac{N_c}{384\pi^2},
~~~~~~~~L_9^{\rm (anom)}
=\frac{N_c}{48\pi^2},~~~~~~~~~~
L_{10}^{\rm (anom)}
=-\frac{N_c}{96\pi^2}
\nonumber,\\
&&H_1^{\rm (anom)}
=\frac{N_c}{96\pi^2}\lim_{\kappa\rightarrow 0}
\lim_{\Lambda\rightarrow\infty}(\ln\frac{\kappa^2}{\Lambda^2}
+\gamma+\frac{1}{2})
~~~~~~~~~~~H_2^{\rm (anom)}
=\frac{N_c}{192\pi^2}+
\lim_{\Lambda\rightarrow\infty}\frac{N_c\Lambda^2}{32\pi^2B_0^2}.
\label{anomalyresults}
\end{eqnarray}
These exactly coincide with the results with $M_Q=0$ in Ref.\cite{Espriu}.
Note that the final expressions of the coefficients $~L_1,\cdots
,L_{10}~$ are independent of the infrared cutoff parameter $\kappa$ and the 
ultraviolet cutoff $\Lambda$ although these cutoff parameters appear in 
${\cal K}_1^{\rm (anom)},\cdots ,{\cal K}_{15}^{\rm (anom)}$, while 
$H_1$ and $H_2$ depend on the cutoff parameters. This implies that $H_1$ 
and $H_2$ are not measurable quantities.
With $N_c=3$, the values of the coefficients are (in units of $10^{-3}$)
\begin{eqnarray}                              
&&L_1=0.79,~~~~~~~~~~L_2=1.58,~~~~~~~~~~L_3=-3.17,~~~~~~~~~~L_4=L_5=L_6=0,
\nonumber\\
&&L_7=0.26,
~~~~~~~~~~L_8=-0.79,~~~~~~~~L_9=6.33,~~~~~~~~~~~~L_{10}=-3.17.
\label{anomalyL}
\end{eqnarray}
These are to be compared with the experimental values (in units of $10^{-3}$)
\cite{GS}
\begin{eqnarray}                               
&&L_1=0.9\pm 0.3,~~~~~~L_2=1.7\pm 0.7,~~~~~L_3=-4.4\pm
2.5,~~~~~L_4=0\pm 0.5,~~~~~L_5=2.2\pm 0.5,~~~~~
L_6=0\pm 0.3,
\nonumber\\
&&L_7=-0.4\pm 0.15,~~L_8=1.1\pm
0.3,~~~~~L_9=7.4\pm 0.7,~~~~~~~~L_{10}=-6.0\pm 0.7.
\label{exptL}
\end{eqnarray}
The numbers in Eqs.(\ref{anomalyL}) are close to the experimental results of
Eqs.(\ref{exptL}) except $L_7$ and $L_8$ are of wrong signs.
This gives people an impression that the coefficients
$~L_1,\cdots ,L_{10}~$ might mainly be contributed by the anomaly part,
and the normal part might only contribute small corrections 
\cite{Espriu,Simic}. However,
we note that the results in Eqs.(\ref{anomalyresults}) are independent of QCD 
interactions, i.e., these terms remain unchanged when we switch off the QCD
gauge coupling constant $\alpha_s$. This is somewhat confusing since these 
coefficients indicate meson interactions which should be residual interactions
between quarks and gluons. We shall see in the next section that {\it these 
terms will actually be completely cancelled by the terms independent of
the quark self-energy in the normal part contribution, so that they do not 
really appear in the final form of the coefficients. What appear in the 
coefficients are the remaining terms in the normal part 
contribution which depend on the quark self-energy and hence on the QCD 
interactions as it should be.} Another feature of the terms in
Eqs.(\ref{anomalyresults}) indicating that they should be exactly cancelled
and should not appear in the final formulae for the coefficients is
the divergence of $H_1$ and $H_2$ when taking $\Lambda\to\infty$. We
know from Ref.\cite{GS} that the ultraviolet divergences in the $O(p^4)$ chiral
Lagrangian coefficients come merely from the meson-loop corrections
with the $O(p^2)$ interactions. In the $1/N_c$ expansion, the 
meson-loop corrections belonging to $O(1/N_c)$ will not take place in the
large-$N_c$ limit. Therefore, in the large-$N_c$ limit, the final expressions 
for the $O(p^4)$ coefficients should be finite when $\Lambda\to\infty$.
Now the ultraviolet divergences in $H_1$ and $H_2$ in 
Eqs.(\ref{anomalyresults}) have nothing to do with the meson-loop corrections, 
so that they should be exactly cancelled by other terms and should not appear 
in the final expressions for the $O(p^4)$ coefficients.

\section{On the Contributions from the Normal part}

In this section we use the same regularization technique as in Sec. II
to calculate the normal part contributions to the chiral Lagrangian
coefficients. We start from the effective action $S_{\rm eff}^{({\rm norm})}$ 
given in Ref.\cite{WKWX1},
\begin{eqnarray}                             
\displaystyle
&&e^{iS_{\rm eff}^{({\rm nrom})}}=\int {\cal D}\Xi 
e^{i\tilde{\Gamma}[1,J_\Omega,\Xi,
\Phi_\Omega,\Pi_\Omega]}
\nonumber\\
&&\hspace{0.4cm}=\int{\cal D}\Xi{\cal D}\Phi_{\Omega}\exp\bigg\{
i\Gamma_0[J_{\Omega},\Phi_{\Omega},\Pi_{\Omega c}]+i\Gamma_I[\Phi_{\Omega}]
+iN_c\int d^4x{\rm tr}_{lf}\{\Xi(x)
[-i\sin\frac{\vartheta(x)}{N_f}
+\gamma_5\cos\frac{\vartheta(x)}{N_f}]
\Phi_{\Omega}^T(x,x)\}\bigg\}
\label{Seff}
\end{eqnarray}
which satisfies a useful relation \cite{WKWX1}
\begin{eqnarray}                           
\frac{d S_{\rm eff}^{({\rm norm})}}
{d J^{\sigma\rho}_{\Omega}(x)}
\bigg|_{U \mbox{ fix, anomaly ignored}}
=N_c\overline{\Phi_{\Omega c}
^{\sigma\rho}(x,x)}\label{Seffdiff}.
\end{eqnarray}
The symbols are defined in Ref.\cite{WKWX1}.

In the large-$N_c$ limit, the integrations in Eq.(\ref{Seff}) can be
carried out by the saddle point approximation with the saddle point equations
\begin{eqnarray}                        
&&\Phi^{(a\eta)(b\zeta)}_{\Omega c}(x,y)=-i[(i\partial\!\!\! /\;+J_{\Omega}
-\Pi_{\Omega c})^{-1}]^{(b\zeta)(a\eta)}(y,x),
\label{Pieq}\\
&&\Pi_{\Omega c}^{\sigma\rho}(x,y)=-\tilde{\Xi}^{\sigma\rho}(x)\delta^4(x-y)
-\sum^{\infty}_{n=1}{\int}d^{4}x_1\cdots{d^4}x_{n}
d^{4}x_{1}'\cdots{d^4}x_{n}'\frac{(-i)^{n+1}(N_c g_s^2)^n}{n!}\nonumber\\
&&\hspace{0.2cm}\times\overline{G}^{\sigma\sigma_1\cdots\sigma_n}_{\rho\rho_1
\cdots\rho_n}(x,y,x_1,x'_1,\cdots,x_n,x'_n)
\Phi_{\Omega c}^{\sigma_1\rho_1}(x_1 ,x'_1)\cdots 
\Phi_{\Omega c}^{\sigma_n\rho_n}(x_n ,x'_n),
\label{fineqNc}\\
&&{\rm tr}_l\bigg[\bigg(-i\sin\frac{\vartheta(x)}{N_f}
+\gamma_5\cos\frac{\vartheta(x)}{N_f}\bigg)\Phi^T_{\Omega c}(x,x)\bigg]=0,
\label{constraint}
\end{eqnarray}
where $\tilde{\Xi}$ is a short notation for the following quantity
\begin{eqnarray}                                   
\tilde{\Xi}^{\sigma\rho}(x)\equiv \frac{\partial}
{\partial\Phi^{\sigma\rho}_{\Omega c}(x,x)}
\int d^4y\; {\rm tr}_{lf}\{\Xi_c(y)[-i\sin\frac{\vartheta_c(y)}{N_f}
+\gamma_5\cos\frac{\vartheta_c(y)}{N_f}]\Phi_{\Omega c}^T(y,y)\}
\bigg|_{\Xi_c\mbox{ fixed}}.\label{xitilde}
\end{eqnarray}
Then the obtained $S_{\rm eff}^{({\rm norm})}$ in this approximation is
\begin{eqnarray}                                
S_{\rm eff}^{({\rm norm})}
&=&\tilde{\Gamma}[1,J_{\Omega},\Xi_c,\Phi_{\Omega c},\Phi_{\Omega c}]
\nonumber\\
&=&-iN_c{\rm Tr}\ln[i\partial\!\!\!/+J_{\Omega}-\Pi_{\Omega c}]
+N_c\int d^{4}xd^{4}x'
\Phi^{\sigma\rho}_{\Omega c}(x,x')\Pi^{\sigma\rho}_{\Omega c}(x,x')
+N_c\sum^{\infty}_{n=2}{\int}d^{4}x_1\cdots d^4x_{n}'\nonumber\\
&&\times
\frac{(-i)^{n}(N_c g_s^2)^{n-1}}{n!}\bar{G}^{\sigma_1\cdots\sigma_n}_{\rho_1
\cdots\rho_n}(x_1,x'_1,\cdots,x_n,x'_n)
\Phi^{\sigma_1\rho_1}_{\Omega c}(x_1 ,x'_1)\cdots 
\Phi^{\sigma_n\rho_n}_{\Omega c}(x_n ,x'_n)\nonumber\\
&&+iN_c \int d^4x~{\rm tr}_{lf}\{\Xi_c(x)[-i\sin\frac{\vartheta_c(x)}{N_f}
+\gamma_5\cos\frac{\vartheta_c(x)}{N_f}]\Phi_{\Omega,c}^T(x,x)\},
\label{Snorm}
\end{eqnarray}
in which the $O(1/N_c)$ term $\Gamma_I$ is neglected. Note that the
last term in Eq.(\ref{Snorm}) actually vanishes due to Eq.(\ref{constraint}).
We keep it here for showing the relation between the effective action
$S_{\rm eff}^{({\rm norm})}$ and its stationary conditions Eqs.(\ref{Pieq})$\--$
(\ref{constraint}).

In the large-$N_c$ limit, $\overline{\Phi_{\Omega c}}=\Phi_{\Omega c}$
on the right hand side of Eq.(\ref{Seffdiff}). The left hand side of
Eq.(\ref{Seffdiff}) can be carried out from Eq.(\ref{Snorm}) using 
Eqs.(\ref{Pieq})$\--$(\ref{constraint}). Then the explicit form of 
Eq.(\ref{Seffdiff}) in this approximation is
\begin{eqnarray}                            
-i[(i\partial\!\!\! /\;+J_{\Omega}
-\Pi_{\Omega c})^{-1}]^{\rho\sigma}(x,x)
=\Phi^{\sigma\rho}_{\Omega c}(x,x).
\label{Eq}
\end{eqnarray}
We see that $\Pi_{\Omega c}$ and $\Phi_{\Omega c}$ play the roles of
the quark self-energy and the quark propagator, respectively, in the
case with $J_\Omega\ne 0$.

Now we decompose $S_{\rm eff}^{({\rm norm})}$ into a part {\it independent of
$\Pi_{\Omega c}$} and a part {\it depending on $\Pi_{\Omega c}$}. The
part independent of $\Pi_{\Omega c}$ can be extracted from
$S_{\rm eff}^{({\rm norm})}$ by setting $\Pi_{\Omega c}=0$, i.e.,
\begin{eqnarray}                            
S^{({\rm norm},\Pi_{\Omega c}=0)}_{\rm eff}
&=&-iN_c{\rm Tr}\ln[i\partial\!\!\!/+J_{\Omega}]
+N_c\bigg[\sum^{\infty}_{n=2}{\int}d^{4}x_1\cdots d^4x_{n}'
\frac{(-i)^{n}(N_c g_s^2)^{n-1}}{n!}\bar{G}^{\sigma_1\cdots\sigma_n}_{\rho_1
\cdots\rho_n}(x_1,x'_1,\cdots,x_n,x'_n)\nonumber\\
&&\times\Phi^{\sigma_1\rho_1}_{\Omega c}(x_1 ,x'_1)\cdots 
\Phi^{\sigma_n\rho_n}_{\Omega c}(x_n ,x'_n)\bigg]_{\Pi_{\Omega c}=0}.
\label{SPi=0}
\end{eqnarray}
Here we have ignored the last term in Eq.(\ref{Snorm}) which actually
vanishes due to Eq.(\ref{constraint}).
We show in APPENDIX B that the last term in Eq.(\ref{SPi=0}) is actually 
$\Omega$-independent. Therefore, Eq.(\ref{SPi=0}) can be written as 
\begin{eqnarray}                            
S^{({\rm norm},\Pi_{\Omega c}=0)}_{\rm eff}
&=&-iN_c[{\rm Tr}\ln(i\partial\!\!\!/+J_{\Omega})
+\Omega-{\rm independent~terms}].
\label{SPi=0'}
\end{eqnarray}
Comparing the $J_\Omega$-dependent terms in Eqs.(\ref{Sanom}) and
(\ref{SPi=0'}), we see that they are of the same form but with an
opposite sign. Thus {\it their contributions to the chiral Lagrangian 
coefficients exactly cancel each other to all orders in the momentum 
expansion}. The cancellation in the case of the $O(p^2)$ coefficient $F^2_0$
has been described in footnote 1 in Sec. II. For the $O(p^4)$ coefficients, 
we have
\begin{eqnarray}                            
&&{\cal K}_i^{(\rm anom)}+{\cal K}_i^{(\Pi_{\Omega c}=0)}=0,~~~~i=1,\cdots,16
\nonumber\\
&&L_i^{(\rm anom)}+L_i^{(\rm norm,\Pi_{\Omega c}=0)}=0,
~~~~i=1,\cdots,10,\nonumber\\
&&H_i^{(\rm anom)}+H_i^{(\rm norm,\Pi_{\Omega c}=0)}=0,~~~~i=1,2,\nonumber\\
\label{cancel}
\end{eqnarray}
Thus, in the large-$N_c$ limit, the anomaly part contributed chiral Lagrangian 
coefficients in Eqs.(\ref{anomalyresults}) do not really appear in the final 
results of the chiral Lagrangian coefficients although their values 
Eqs.(\ref{anomalyL}) are close to the experimental values. {\it The chiral 
Lagrangian coefficients are actually contributed from the $\Pi_{\Omega c}\ne 
0$ part of $S_{\rm eff}^{({\rm norm})}$},
\begin{eqnarray}                             
S^{({\rm norm},\Pi_{\Omega c}\ne 0)}_{\rm eff}\equiv 
S_{\rm eff}^{({\rm norm})}-S^{({\rm norm},\Pi_{\Omega c}=0)}_{\rm eff}
\label{SPi}
\end{eqnarray}
{\it which leads to the $\Pi_{\Omega c}\ne 0$ part of ${\cal
K}^{({\rm norm})}_1,\cdots,{\cal K}^{({\rm norm})}_{15}$},
\begin{eqnarray}                             
{\cal K}^{({\rm norm},\Pi_{\Omega c}\ne 0)}_i={\cal K}^{({\rm norm})}_i
-{\cal K}^{({\rm norm},\Pi_{\Omega c}=0)}_i,~~~~~~i=1,\cdots,15.
\label{KPi}
\end{eqnarray}
{\it This is our first new conclusion in this study}. 

The final chiral Lagrangian coefficients are then
\begin{eqnarray}                       
L_i=L_i^{({\rm norm},\Pi_{\Omega c}\ne 0 )},~~~~i=1,\cdots,10,
~~~~~~~~~~~~
H_i=H_i^{({\rm norm},\Pi_{\Omega c}\ne 0)},~~~~i=1,2,
\label{totalL}
\end{eqnarray}
and
\begin{eqnarray}                          
&&L_1
=\frac{1}{32}{\cal K}_4^{^^{^{({\rm norm},\Pi_{\Omega c}\ne 0)}}
+\frac{1}{16}{\cal K}_5^{({\rm norm},\Pi_{\Omega c}\ne 0)}
+\frac{1}{16}{\cal K}_{13}^{^{({\rm norm},\Pi_{\Omega c}\ne 0)}}
-\frac{1}{32}{\cal K}_{14}^{({\rm norm},\Pi_{\Omega c}\ne 0)}\nonumber,\\
&&L_2
=\frac{1}{16}({\cal K}_4^{({\rm norm},\Pi_{\Omega c}\ne 0)}
+{\cal K}_6^{({\rm norm},\Pi_{\Omega c}\ne 0)})+\frac{1}{8}
{\cal K}_{13}^{({\rm norm},\Pi_{\Omega c}\ne 0)}
-\frac{1}{16}{\cal K}_{14}^{({\rm norm},\Pi_{\Omega c}\ne 0)}\nonumber,\\
&&L_3
=\frac{1}{16}({\cal K}_3^{({\rm norm},\Pi_{\Omega c}\ne 0)}
-2{\cal K}_4^{({\rm norm},\Pi_{\Omega c}\ne 0)}
-6{\cal K}_{13}^{({\rm norm},\Pi_{\Omega c}\ne 0)}
+3{\cal K}_{14}^{({\rm norm},\Pi_{\Omega c}\ne 0)})\nonumber,\\
&&L_4
=\frac{{\cal K}_{12}^{({\rm norm},\Pi_{\Omega c}\ne 0)}}{16B_0}\nonumber,~~
L_5
=\frac{{\cal K}_{11}^{({\rm norm},\Pi_{\Omega c}\ne 0)}}{16B_0}\nonumber,~~
L_6
=\frac{{\cal K}_8^{({\rm norm},\Pi_{\Omega c}\ne 0)}}{16B_0^2}\nonumber,\\
&&L_7
=-\frac{{\cal K}_1^{({\rm norm},\Pi_{\Omega c}\ne 0)}}{16N_f}
-\frac{{\cal K}_{10}^{({\rm norm},\Pi_{\Omega c}\ne 0)}}{16B_0^2}
-\frac{{\cal K}_{15}^{({\rm norm},\Pi_{\Omega c}\ne 0)}}{16B_0N_f}\nonumber,\\
&&L_8
=\frac{1}{16}[{\cal K}_1^{({\rm norm},\Pi_{\Omega c}\ne 0)}
+\frac{1}{B_0^2}{\cal K}_7^{({\rm norm},\Pi_{\Omega c}\ne 0)}
-\frac{1}{B_0^2}{\cal K}_9^{({\rm norm},\Pi_{\Omega c}\ne 0)}
+\frac{1}{B_0}{\cal K}_{15}^{({\rm norm},\Pi_{\Omega c}\ne 0)}]\nonumber,\\
&&L_9
= \frac{1}{8}(4{\cal K}_{13}^{({\rm norm},\Pi_{\Omega c}\ne 0)}
-{\cal K}_{14}^{({\rm norm},\Pi_{\Omega c}\ne 0)})\nonumber,\\
&&L_{10}
=\frac{1}{2}({\cal K}_2^{({\rm norm},\Pi_{\Omega c}\ne 0)}
-{\cal K}_{13}^{({\rm norm},\Pi_{\Omega c}\ne 0)})\nonumber,\\
&&H_1
=-\frac{1}{4}({\cal K}_2^{({\rm norm},\Pi_{\Omega c}\ne 0)}
+{\cal K}_{13}^{({\rm norm},\Pi_{\Omega c}\ne 0)})\nonumber,\\
&&H_2
=\frac{1}{8}[-{\cal K}_1^{({\rm norm},\Pi_{\Omega c}\ne 0)}
+\frac{1}{B_0^2}{\cal K}_7^{({\rm norm},\Pi_{\Omega c}\ne 0)}
+\frac{1}{B_0^2}{\cal K}_9^{({\rm norm},\Pi_{\Omega c}\ne 0)}
-\frac{1}{B_0}{\cal K}_{15}^{({\rm norm},\Pi_{\Omega c}\ne 0)}].
\label{p4full}
\end{eqnarray}
Since $\Pi_{\Omega c}|_{g_s=0}=0$, these $O(p^4)$ chiral Lagrangian
coefficients will vanish if we switch off the QCD coupling constant $g_s$
as it should be.

\section{Calculation of the Chiral Lagrangian Coefficients}

We see that to calculate the chiral Lagrangian coefficients from
$S_{\rm eff}^{({\rm norm},\Pi_{\Omega c}\ne 0)}$, we should mainly deal
with $S_{\rm eff}^{({\rm norm})}$ given in Eq.(\ref{Snorm}) which has
never been carefully calculated in the literature.
Ignoring the vanishing last term in Eq.(\ref{Snorm}), there are still
rather complicated terms in it. For example, the third term includes
various ranks of gluon Green's functions which concern very complicated
calculations of QCD dynamics.  
As the first time of doing this kind of calculation, we shall take
futher approximations to simplify the evaluation of 
$S_{\rm eff}^{({\rm norm})}$. 
We know that the pion decay constant $f_\pi$ has been studied from QCD in 
Ref.\cite{PS} by taking the approximation of keeping only the leading order in 
dynamical perturbation, i.e., taking into account only the QCD interaction in 
the Schwinger-Dyson equation leading to the nonperturbative solution of chiral 
symmetry breaking, and neglecting other QCD corrections in positive powers of 
$g_s$ (perturbative). This approximation leads to the widely used 
Pagels-Stokar formula which is reasonable though not perfect.
In the large-$N_c$ limit, $f_\pi$ is just the $O(p^2)$ chiral Lagrangian 
coefficient $F_0$ given in Eqs.(\ref{F_0B_0}). So, as in Ref.\cite{PS}, we 
take the approximation of keeping only the leading order in dynamical
perturbation to calculate the chiral Lagrangian coefficients from 
$S_{\rm eff}^{({\rm norm})}$. In this spirit, we neglect the
complicated third term in Eq.(\ref{Snorm}) which contains only positive
powers of $g_s$. Furthermore, we see from Eq.(\ref{fineqNc}) that the second
term in Eq.(\ref{Snorm}) is of the same order as the third term, so that
we neglect the second term in Eq.(\ref{Snorm}) as well. With this
approximation, $S_{\rm eff}^{({\rm norm})}$ is simplified as
\begin{eqnarray}                         
S_{\rm eff}^{({\rm norm})}
=-iN_c{\rm Tr}\ln[i\partial\!\!\!/+J_{\Omega}-\Pi_{\Omega c}].
\label{Snormsimple}
\end{eqnarray}
Now the concerned QCD dynamics resides in the $\Pi_{\Omega c}$ term which is 
related to the quark self-energy [cf. Eq.(\ref{Eq})]. We expect such a
simple approximation may also lead to reasonable results of the $O(p^4)$
chiral Lagrangian coefficients since reasonable values of the $O(p^4)$ chiral 
Lagrangian coefficients have been obtained in a model by Holdom
\cite{Holdom} considering only the quark self-energy
contribution with certain phenomenological ansatz. We shall see in Sec. V that 
our obtained $O(p^4)$ chiral Lagrangian coefficients are indeed reasonable.
Although this approximation is crude, it provides a simple illustration
of the main feature of how QCD predicts the chiral Lagrangian coefficients.
Further improved study beyond this simple approximation
taking into account the second and third terms in Eq.(\ref{Snorm}) is
of course needed. That will be presented in a later paper.
Now we need to calculate $\Pi^{\sigma\rho}_{\Omega c}(x,y)$ and carry out
the explicit expression for $S_{\rm eff}^{({\rm norm})}$ in 
Eq.(\ref{Snormsimple}).

We have noticed that $\Pi^{\sigma\rho}_{\Omega c}$ is related to the
quark self-energy. If we find out the relation between 
$\Pi^{\sigma\rho}_{\Omega c}(x,y)$ and the conventional quark self-energy 
$\Sigma(-p^2)$, then we can obtain $\Sigma(-p^2)$ by solving the well-known 
Schwinger-Dyson equation. As in the literature, we shall write down the
Schwinger-Dyson equation in the Landau gauge which is stable against the
gauge parameter. In the same approximation of taking 
the leading order in dynamical perturbation theory and with the improved
ladder approximation, the Schwinger-Dyson equation in the Euclidean 
space-time reads \cite{Aoki,Munczek,Dai}
\begin{eqnarray}                      
\Sigma(p^2)-\frac{3N_c}{2}\int\frac{d^4q}{4\pi^3}
\frac{\alpha_s[(p-q)]}{(p-q)^2}
\frac{\Sigma(q^2)}{q^2+\Sigma^2(q^2)}
=0.
\label{eq0}
\end{eqnarray}
This equation can be solved numerically, and the
details will be presented in Sec. V. Naively, we may expect that 
$\Pi^{\sigma\rho}_{\Omega c}(x,y)=\delta^{\sigma\rho}\Sigma(\partial^2_x)
\delta^4(x-y)$. But this is not correct. Under a local chiral
transformation $h(x)$ (hidden symmetry transformation \cite{WKWX1}), 
$\Pi^{\sigma\rho}_{\Omega c}$ transforms as
\begin{eqnarray}                         
\Pi_{\Omega c}(x,y)\rightarrow\Pi_{\Omega c}'(x,y)=h^{\dagger}(x)
\Pi_{\Omega c}(x,y)h(y),
\label{hidden0}
\end{eqnarray}
while $\delta^{\sigma\rho}\Sigma(\partial^2_x)\delta^4(x-y)$ does not
transform like this. The correct relation can be found by replacing
the ordinary derivative $\partial_x^\mu$ by the covariant derivative
\begin{eqnarray}                      
\overline{\nabla}^{\mu}_x=\partial^{\mu}_x-iv_{\Omega}^{\mu}(x).
\end{eqnarray}
Since the external source $v^\mu_\Omega(x)$ transforms as
\begin{eqnarray}                     
v_{\Omega}^{\mu}(x)&\rightarrow&v_{\Omega}^{\mu\prime}(x)=h^{\dagger}(x)
v_{\Omega}^{\mu}(x)h(x)+ih^{\dagger}(x)[\partial^{\mu}h(x)],
\end{eqnarray}
the covariant derivative $\overline{\nabla}^\mu_x$ transforms as
\begin{eqnarray}                       
\overline{\nabla}_x^{\mu}&\rightarrow&\overline{\nabla}_x^{\mu\prime}
=h^{\dagger}(x)
\overline{\nabla}_x^{\mu}h(x).
\end{eqnarray}
Thus the correct identification is
\begin{eqnarray}                      
\Pi^{\sigma\rho}_{\Omega c}(x,y)=
[\Sigma(\overline{\nabla}^2_x)]^{\sigma\rho}\delta^4(x-y). 
\end{eqnarray}
Then the effective action (\ref{Snormsimple}) can be written as
\begin{eqnarray}                      
S_{\rm eff}^{({\rm norm})}
=-iN_c{\rm Tr}\ln[i\partial\!\!\!/+J_{\Omega}
-\Sigma(\overline{\nabla}^2)].
\label{SGL}
\end{eqnarray}

Next we evaluate the effective action (\ref{SGL}) using the Schwinger
proper time regularization as before [cf. APPENDIX A] to obtain the
expressions for the chiral Lagrangian coefficients. This is not trivial
since usually this regularization scheme is used in the case that
$\Sigma$ is a constant, and thus the momentum integration can be explicitly
carried out to check the local gauge invariance of the result. Now we leave
$\Sigma(\overline{\nabla}^2)$ as an unspecified function in Eq.(\ref{SGL}), so 
that the momentum integration cannot be carried out explicitly. Organizing 
terms to guarantee local chiral invariance is rather tedious and the details
are given in Ref.\cite{det}. Our obtained expressions in the Minkowskian
space-time are
\begin{eqnarray}             
&&F_0^2B_0=4\int d\tilde{p} \Sigma_pX_p,
\label{F0B0}\\
&&F_0^2=2\int d\tilde{p}\bigg[(-2\Sigma^2_p-p^2\Sigma_p\Sigma'_p)X_p^2
+(2\Sigma^2_p+p^2\Sigma_p\Sigma'_p)\frac{X_p}{\Lambda^2}\bigg],
\label{F0}\\
&&{\cal K}^{({\rm norm})}_{1}=2\int d\tilde{p}\bigg[-2A_pX_p^3
+2A_p\frac{X_p^2}{\Lambda^2}-A_p\frac{X_p}{\Lambda^4}
+\frac{p^2}{2}\Sigma'^2_p\frac{X_p}{\Lambda^2}-\frac{p^2}{2}\Sigma'^2_pX_p^2,
\bigg],
\nonumber\\
&&{\cal K}^{({\rm norm})}_{2}=\int d\tilde{p}\bigg[-2B_pX_p^3
 +2B_p\frac{X_p^2}{\Lambda^2}-B_p\frac{X_p}{\Lambda^4}
+\frac{p^2}{2}\Sigma^{\prime 2}_p\frac{X_p}{\Lambda^2},
-\frac{p^2}{2}\Sigma^{\prime 2}_pX_p^2\bigg],
\nonumber\\
&&{\cal K}^{({\rm norm})}_{3}=2\int d\tilde{p}\bigg[
(\frac{4\Sigma^4_p}{3}-\frac{2p^2\Sigma^2_p}{3}+\frac{p^4}{18})(
 6X_p^4-\frac{6X_p^3}{\Lambda^2}+\frac{3X_p^2}{\Lambda^4}
-\frac{X_p}{\Lambda^6}),
\nonumber\\
&&\hspace{0.4cm}+(-4\Sigma^2_p+\frac{p^2}{2})(-2X_p^3
+\frac{2X_p^2}{\Lambda^2}-\frac{X_p}{\Lambda^4})
-\frac{X_p}{\Lambda^2}+X_p^2\bigg],
\nonumber\\
&&{\cal K}^{({\rm norm})}_{4}=\int d\tilde{p}\bigg[
(\frac{-4\Sigma^4_p}{3}+\frac{2p^2\Sigma^2_p}{3}+\frac{p^4}{18})(
6X_p^4-\frac{6X_p^3}{\Lambda^2}+\frac{3X_p^2}{\Lambda^4}
-\frac{X_p}{\Lambda^6})
+4\Sigma^2_p(-2X_p^3+\frac{2X_p^2}{\Lambda^2}
\nonumber\\
&&\hspace{0.4cm}-\frac{X_p}{\Lambda^4})+\frac{X_p}{\Lambda^2}-X_p^2\bigg],
\nonumber\\
&&{\cal K}^{({\rm norm})}_5={\cal K}^{({\rm norm})}_6=0,
\nonumber\\
&&{\cal K}^{({\rm norm})}_7=2\int d\tilde{p}\bigg[(3\Sigma^2_p+2p^2\Sigma_p\Sigma'_p)X_p^2
+[-2\Sigma^2_p-p^2(1+2\Sigma_p\Sigma'_p)]\frac{X_p}{\Lambda^2}\bigg],
\nonumber\\
&&{\cal K}^{({\rm norm})}_8=0,
\nonumber\\
&&{\cal K}^{({\rm norm})}_9=2\int d\tilde{p}\bigg[(\Sigma^2_p+2p^2\Sigma_p\Sigma'_p)X_p^2
-p^2(1+2\Sigma_p\Sigma'_p)\frac{X_p}{\Lambda^2}\bigg],
\nonumber\\
&&{\cal K}^{({\rm norm})}_{10}=0,
\nonumber\\
&&{\cal K}^{({\rm norm})}_{11}=4\int d\tilde{p}
\bigg[(-4\Sigma^3_p+p^2\Sigma_p)X_p^3
+(4\Sigma^3_p-p^2\Sigma_p)\frac{X_p^2}{\Lambda^2}
-(2\Sigma^3_p-\frac{1}{2}p^2\Sigma_p)\frac{X_p}{\Lambda^4}
+3\Sigma_p\frac{X_p}{\Lambda^2}
\nonumber\\
&&\hspace{0.4cm}-3\Sigma_p X_p^2\bigg],
\nonumber\\
&&{\cal K}^{({\rm norm})}_{12}=0,
\nonumber\\
&&{\cal K}^{({\rm norm})}_{13}=\int d\tilde{p}\bigg[
(\frac{1}{3}p^2\Sigma'_p\Sigma''_p+\frac{1}{3}\Sigma_p\Sigma''_p)X_p
+(C_p-D_p)\frac{X_p}{\Lambda^2}
-(C_p-D_p)X_p^2-2E_pX_p^3
\nonumber\\
&&\hspace{0.4cm}+2E_p\frac{X_p^2}{\Lambda^2}
-E_p\frac{X_p}{\Lambda^4}\bigg],
\nonumber\\
&&{\cal K}^{({\rm norm})}_{14}=-4\int d\tilde{p}\bigg[
-2F_pX_p^3+2F_p\frac{X_p^2}{\Lambda^2}
-F_p\frac{X_p}{\Lambda^4}
+\frac{p^2}{2}\Sigma_p^{\prime 2}\frac{X_p}{\Lambda^2}
-\frac{p^2}{2}\Sigma^{\prime 2}_pX_p^2\bigg],
\nonumber\\
&&{\cal K}^{({\rm norm})}_{15}=-4\int d\tilde{p}\bigg[
-(\Sigma_p+\frac{1}{2}p^2\Sigma'_p)\frac{X_p}{\Lambda^2}
+(\Sigma_p+\frac{1}{2}p^2\Sigma'_p)X_p^2\bigg],
\label{Kresult}
\end{eqnarray}
in which the short notations (in the Minkowskian space-time) are
\begin{eqnarray}                     
&&\Sigma_p\equiv\Sigma(-p^2),\\
&&\int d\tilde{p}\equiv
iN_c\int\frac{d^4p}{(2\pi)^4}e^{\frac{p^2-\Sigma^2_p}{\Lambda^2}},
\label{measure}
\nonumber\\
&&X_p\equiv\frac{1}{p^2-\Sigma^2_p},
\nonumber\\
&&A_p=-\frac{2}{3}p^2\Sigma_p\Sigma'_p(-1-2\Sigma_p\Sigma'_p)-\frac{1}{3}
\Sigma^2_p(-1-2\Sigma_p\Sigma'_p)
+\frac{1}{3}p^2\Sigma^2_p(-\Sigma^{\prime 2}_p-
\Sigma_p\Sigma''_p)
\nonumber\\ 
&&\hspace{0.4cm}-\frac{1}{6}p^4(-\Sigma^{\prime
2}_p -\Sigma_p\Sigma''_p),
\nonumber\\
&&B_p=-\frac{2}{3}p^2\Sigma_p\Sigma'_p(-1-2\Sigma_p\Sigma'_p)-\frac{1}{3}
\Sigma^2_p(-1-2\Sigma_p\Sigma'_p)+\frac{1}{3}p^2\Sigma^2_p(-\Sigma^{\prime
2}_p -\Sigma_p\Sigma''_p)
\nonumber\\
&&\hspace{0.4cm}-\frac{1}{18}p^4(-\Sigma^{\prime 2}_p-\Sigma_p\Sigma''_p)
-\frac{1}{6}p^2(-1-2\Sigma_p\Sigma'_p),
\nonumber\\
&&C_p=\frac{1}{3}-\frac{1}{3}\Sigma_p\Sigma'_p
-\frac{1}{2}p^2\Sigma^{\prime 2}_p,
\nonumber\\
&&D_p=\frac{1}{2}p^2\Sigma^{\prime
2}_p-\frac{1}{6}p^2\Sigma_p\Sigma''_p
(-1-2\Sigma_p\Sigma'_p)-\frac{2}{9}p^4\Sigma'_p\Sigma''_p
(-1-2\Sigma_p\Sigma'_p) -\frac{2}{9}p^4\Sigma_p^{\prime
2}(-\Sigma^{\prime 2}_p-\Sigma_p\Sigma''_p) 
\nonumber\\
&&\hspace{0.4cm}-\frac{1}{3}p^2\Sigma_p\Sigma'_p(-\Sigma^{\prime
2}_p-\Sigma_p\Sigma^{\prime\prime}_p),
\nonumber\\
&&E_p=-\frac{1}{6}p^2\Sigma_p\Sigma'_p(-1-2\Sigma_p\Sigma'_p)^2
-\frac{1}{9}
p^4\Sigma^{\prime
2}_p(-1-2\Sigma_p\Sigma'_p)^2,
\nonumber\\
&&F_p=-\frac{4}{3}p^2\Sigma_p\Sigma'_p+
\frac{4}{3}p^2(\Sigma_p\Sigma'_p)^2-\frac{2}{3}\Sigma^2_p
+\frac{2}{3}\Sigma_p^3\Sigma'_p
+\frac{1}{3}p^2\Sigma^2_p(-\Sigma^{\prime
2}_p-\Sigma_p\Sigma''_p)
\nonumber\\
&&\hspace{0.4cm}-\frac{1}{9}p^4(-\Sigma^{\prime 2}_p-\Sigma_p\Sigma''_p)
-\frac{1}{3}p^2(-1-2\Sigma_p\Sigma'_p)-\frac{1}{2}p^2.
\end{eqnarray}

For the coefficient $F_0^2$, Eq.(\ref{F0}) is just the well-known 
Pagels-Stokar formula \cite{PS} when taking the regularization cutoff 
parameter $\Lambda\to \infty$. 

It is easy to check that these ${\cal K}^{({\rm norm})}_i$ ($i=1,\cdots,15$) 
do contain the $\Pi_{\Omega c}$-independent ($\Sigma_p$-independent) piece 
which exactly cancel the anomaly contributions in Eqs.(\ref{anomK}) mentioned 
in Sec. III. This can be done by taking a constant $\Sigma_p$ to carry out the 
momentum integrations, and picking up the $\Sigma_p$-independent terms which 
are just the $\Pi_{\Omega c}$-independent terms mentioned in Sec. III.
Subtracting these $\Pi_{\Omega c}$-independent terms from the obtained
${\cal K}^{({\rm norm})}_i$ in Eqs.(\ref{Kresult}), we get the desired
${\cal K}^{({\rm norm},\Pi_{\Omega c}\ne 0)}_i$ in Eq.(\ref{KPi}), which
is needed in the final expressions for the chiral Lagrangian coefficients 
in Eqs.(\ref{p4full}). 

We can also check that the regularization cutoff $\Lambda$ does not appear in
${\cal K}^{({\rm norm},\Pi_{\Omega c}\ne 0)}_i$, so that the obtained
chiral Lagrangian coefficients $~L_1,\cdots,~L_{10}~$ are all finite as it 
should be since there is no divergence in the large-$N_c$ limit [the divergent 
meson-loop corrections are of $O(1/N_c)$].

\section{Numerical Calculations}

The last step in the calculation is to solve the Schwinger-Dyson
equation (\ref{eq0}) numerically to obtain $\Sigma(p^2)$. In the
integrand in the Schwinger-Dyson equation (\ref{eq0}), there is still the QCD 
running coupling constant $\alpha_s(p-q)$ unspecified. The high momentum 
behavior of $\alpha_s$ is well known. The one-loop level formula is
\begin{eqnarray}                        
\alpha_s(p)\stackrel{p^2\to\infty}\longrightarrow
\frac{12\pi}{(33-2N_f)}\frac{1}{\ln(p^2/\Lambda^2_{QCD})}.
\label {alphasUV}
\end{eqnarray}
The low momentum behavior of $\alpha_s(p)$ is not known yet due to the
ignorance of nonperturbative QCD. Inevitably, we have to take certain
QCD motivated model for it as in the literature. We shall take the following 
Model A from Ref.\cite{Aoki}, and Model B and Model C from Ref.\cite{Munczek} 
as examples to do the calculation. 
They are
\begin{eqnarray}                      
{\rm A}:~~~~~~~~~~\alpha_s(p)&&=7\frac{12\pi}{(33-2N_f)},
\hspace{5.22cm} \mbox{for}~\ln(p^2/\Lambda_{QCD}^2)\leq -2;\nonumber\\
&&=\{7-\frac{4}{5}[2+\ln(p^2/\Lambda_{QCD}^2)]^2\}\frac{12\pi}{(33-2N_f)},
\hspace{1.2cm} \mbox{for}~-2\leq\ln(p^2/\Lambda_{QCD}^2)\leq 0.5;\nonumber\\
&&=\displaystyle\frac{1}{\ln(p^2/\Lambda^2_{QCD})}\frac{12\pi}{(33-2N_f)},
\hspace{3.35cm} \mbox{for}~0.5\leq\ln(p^2/\Lambda_{QCD}^2).
\label{omega1}\\
{\rm B}:~~~~~~~~~~\alpha_s(p)&&=  4\pi^3\eta^2p^2
\delta^4(p)+\frac{12\pi}{(33-2N_f)}
\frac{1}{\ln(2+p^2/\Lambda^2_{QCD})};
\label{omega2}\\
{\rm C}:~~~~~~~~~~\alpha_s(p)&&= 
\frac{4\pi^3}{\mu^2}p^2e^{-p^2/p_0^2}+
\frac{12\pi}{(33-2N_f)}\frac{1}{\ln(2+p^2/\Lambda^2_{QCD})}.
\label{omega3}
\end{eqnarray}
They all have the asymptotic behavior (\ref{alphasUV}).
In Eq.(\ref{omega1}), there is only one parameter $\Lambda_{QCD}$, while in 
Eqs.(\ref{omega2}) and (\ref{omega3}), in addition to 
$\Lambda_{QCD}$, there are extra parameters $\eta,~\mu$, and $p_0$, 
respectively. We shall determine the parameters in the following way. 
In the present approach, there are no meson-loop corrections. Thus we should 
identify $F_0=f_\pi=93$ MeV \cite{GS}, and $F_0$ is given by the 
Pagels-Stokar formula. Changing the parameters will cause a change in 
$\Sigma(p^2)$, and thus a change in $F_0$. We take $F_0=93$ MeV as a 
requirement to determine the parameters. In the case of Model A, the 
determined $\Lambda_{QCD}$ is $\Lambda_{QCD}=484$ MeV \cite{Aoki}. In the 
cases of Model B and Model C, there are extra parameters. We take 
the original values $p_0=380$ MeV, and 
$\Lambda_{QCD}=230$ MeV as in Ref.\cite{Munczek}, and determine $\eta$ and 
$\mu$ in the above way. The determined values are $\eta=290$ MeV, and 
$\mu=1160$ MeV~\footnote{The original values of $\eta$ and $\mu$ in 
Ref.\cite{Munczek} are $\eta=920$ MeV, $\mu=600$ MeV which are different from 
ours. The reason is that in Ref.\cite{Munczek} the number of quark flavors is 
taken as $N_f=6$ rather than $N_f=3$, and the formula for $f_\pi$ is more 
complicated than the Pagels-Stokar formula.}. ,
The running coupling constant 
$\alpha_s(p)$ in the three cases are plotted in FIG. 1. We see that they are 
different mainly in the low momentum region.  
 
To solve the Schwinger-Dyson equation (\ref{eq0}), we further take the usual
approximation 
$\alpha_s(p-q)\approx \theta(p^2-q^2)\alpha_s(p^2)+\theta(q^2-p^2)
\alpha_s(q^2)$ 
\cite{angularapprox} with which the angular integration can be easily
carried out, and the integral equation can be converted into the
following differential equation \footnote{In the case of Model B,
there is a term containing $\delta^4(p)$ which is not a function of
$p^2$, and the integration can be directly carried out. Therefore, in
this case, the differential equation and boundary conditions are
different from Eqs.(\ref{fineq1}), (\ref{F01}), and (\ref{Fp01}). They are
$\frac{d}{d p^2}\frac{\frac{d}{d p^2}
\Sigma(p^2)[1-\frac{3N_c/2}{p^2+\Sigma^2(p^2)}]}
{\frac{d}{d p^2}\frac{\beta}{p^2ln(2+p^2/\Lambda^2_{QCD})}}
-{\frac{3N_c}{8\pi}}p^2\frac{\Sigma(p^2)}{p^2+\Sigma^2(p^2)}=0,~~~~~~~~~~~~~~~~~~~~~~~~~~~
\Sigma(\bar\Lambda^2)[1-\frac{3N_c/2}{\bar\Lambda^2+\Sigma(\bar\Lambda^2)}]
-\frac{3N_c}{8\pi}
\frac{\alpha_s(\bar\Lambda^2)}{\bar\Lambda^2}
\int^{\bar\Lambda^2}_0dq^2\frac{q^2\Sigma(q^2)}{q^2+\Sigma(q^2)}=0,
~~
\mbox{and}\\\bigg[\frac{d}{dp^2}(\Sigma(p^2)[1-\frac{3N_c/2}{p^2
+\Sigma(p^2)}])\bigg]_{p^2=0}
+\frac{\frac{3N_c}{16\pi}\alpha_s(0)}{\Sigma(0)}=0.$, respectively.}

\begin{eqnarray}                     
\frac{d}{d p^2}\frac{\Sigma'(p^2)}{\bigg(\frac{\alpha_s(p)}{p^2}\bigg)'}
-{\frac{3N_c}{8\pi}}p^2\frac{\Sigma(p^2)}{p^2+\Sigma^2(p^2)}=0,
\label{fineq1}
\end{eqnarray}
with boundary conditions:
\begin{eqnarray}                     
&&\Sigma(\bar\Lambda^2)-\frac{3N_c}{8\pi}
\frac{\alpha(\bar\Lambda^2)}{\bar\Lambda^2}
\int_0^{\bar\Lambda^2}dq^2\;
\frac{q^2\Sigma(q^2)}{q^2+\Sigma^2(q^2)}=0,
\label{F01}\\
&&\Sigma'(0)+\frac{\frac{3N_c}{16\pi}\alpha_s(0)}{\Sigma(0)}=0,
\label{Fp01}
\end{eqnarray}
where
$\bar\Lambda$ is a momentum cutoff regularizing the integral. We shall
eventually take $\bar\Lambda\to\infty$.

We know that the asymptotic behavior of $\Sigma(p^2)$ reflecting chiral
symmetry breaking is
\begin{eqnarray}                         
&&\Sigma(p^2)\stackrel{p^2\to\infty}\longrightarrow
\frac{\ln^{\gamma-1}(p^2/\Lambda^2_{QCD})}{p^2},
\end{eqnarray}
where $\gamma\equiv (9N_c)/(2(33-2N_f))$.
We have found the numerical solution of Eqs.(\ref{fineq1}), (\ref{F01}) and
(\ref{Fp01}) satisfying this asymptotic behavior. The obtained solution
with $\bar\Lambda\to\infty$ (a large enough number which can be
regarded as infinity) in the three cases are plotted in FIG. 2. Again 
they are different mainly in the low momentum region.

With the obtained $\Sigma(p^2)$, we can calculate the $O(p^4)$ chiral
Lagrangian coefficients from Eqs.(\ref{p4full}), (\ref{KPi}), and (\ref{Kresult}).
The obtained values of $L_1,\cdots,L_{10}$ are listed in TABLE I
together with the experimental values \cite{GS} for comparison. Note
that there is no running of $L_1,\cdots,L_{10}$ in this simple approach
since the meson-loop effects causing the running of $L_1,\cdots,L_{10}$
\cite{GS} are of the order of $1/N_c$, and are neglected in this
approach. Thus the predicted numbers of $L_1,\cdots,L_{10}$ can be
directly compared with the experimental values.
We see from TABLE I that:
\begin{description}
\item{(i)} these coefficients are not so sensitive
to the forms of $\alpha_s(p)$; 
\item{(ii)} all the obtained $L_1,\cdots,L_{10}$ are
of the right orders of magnitude and the right signs; 
\item{(iii)}
$L_1,~L_2,~L_4,~L_6,$ and $L_{10}$ are consistent with the experiments
at the $1\sigma$ level;
\item{(iv)} $L_3,~L_5,~L_7$ and $L_8$ are consistent with
the experiments at the $2\sigma$ level; and
\item{(v)} only $L_9$ deviates from
the experimental value by $(3\--4)\sigma$. 
\end{description}
Considering the large theoretical
uncertainty in this simple approach, {\it the obtained
$L_1,\cdots,L_{10}$ are consistent with the experiments}. We see that
the nonperturbative quark self-energy plays an important role in QCD
contributions to the chiral Lagrangian coefficients. This supports the
phenomenological model of Holdom \cite{Holdom}.

In addition to $L_1,\cdots,L_{10}$, we can also calculate the quark condensate
$\langle\bar\psi\psi\rangle$ from the $O(p^2)$ coefficient $F^2_0B_0$
in Eq.(\ref{F0B0}). In the simple approach in this paper, the relation
between $\langle\bar\psi\psi\rangle$ and $F^2_0B_0$ is \cite{WKWX1}
\begin{eqnarray}                 
\langle\overline{\psi}\psi\rangle=-N_fF_0^2B_0.
\label{condensate}
\end{eqnarray}
We know that, in this simple approach, $F_0=f_\pi=93$ MeV is finite.
But $F^2_0B_0$ in Eq.(\ref{F0B0}) is divergent,
\begin{eqnarray}                  
F_0^2B_0(\Lambda^2,\Lambda^2_{QCD})
\propto\ln^{\gamma}\bigg(\frac{\Lambda^2}{\Lambda^2_{QCD}}\bigg),
\label{div}
\end{eqnarray}
so that it needs to be renormalized. We take a simple renormalization
scheme by taking the counter term as
$F^2_0B_0(\Lambda^2,\mu^2)$, in which $\mu$ is
the renormalization scale~\footnote{This
corresponds to the modified minimal subtraction ($\overline{MS}$) scheme 
\cite{HK}.}. Thus the
renormalized quantity
\begin{eqnarray}               
F^2_0B_{0r}\propto \ln^\gamma\bigg(\frac{\mu^2}{\Lambda^2_{QCD}}\bigg).
\label{renormalization}
\end{eqnarray}
Then the
renormalized (\ref{condensate}) is
\begin{eqnarray}                   
\langle\overline{\psi}\psi\rangle_r=-N_fF_0^2B_{0r}.
\label{remcondensate}
\end{eqnarray}
We take the renormalization scale to be $\mu=1$ GeV to define the quark
condensate. the obtained values of $\langle\bar\psi\psi\rangle_r$ for
the three forms of $\alpha_s(p)$ are
\begin{eqnarray}              
&&{\rm A}:~~~~~~~~~~\langle\bar\psi\psi\rangle_r=-(296~\mbox{MeV})^3,
\nonumber\\
&&{\rm B}:~~~~~~~~~~\langle\bar\psi\psi\rangle_r=-(296~\mbox{MeV})^3,
\nonumber\\
&&{\rm C}:~~~~~~~~~~\langle\bar\psi\psi\rangle_r=-(301~\mbox{MeV})^3.
\label{condensaterenorm}
\end{eqnarray}
These are to be compared with the experimentally determined value 
$\langle\bar\psi\psi\rangle_{expt}=-(250~\mbox{MeV})^3$ from the QCD
sum rule at the scale of the typical hadronic mass \cite{sumrule}. Considering 
the large theoretical uncertainty in this calculation, the predicted quark 
condensate is also consistent with the experiment.

The above results show that {\it the
present simple approach does reveal the main feature of the QCD
predictions for the chiral Lagrangian coefficients} although the
approximations in this approach are rather crude. Of course, further
improvements of the approximations beyond this simple approach are
needed. This kind of study is in progress.

Finally, we would like to mention that, in our calculation, we have taken the 
ultraviolet cutoff parameters $\Lambda,\bar\Lambda\to\infty$, i.e., we
have taken account of the QCD contributions in the whole momentum range.
Note that this has nothing to do with the validity range of the chiral 
Lagrangian determined by the range in which the expansion in the meson 
momentum makes sense, i.e., up to $\Lambda_\chi\approx 4\pi f_\pi$. To see the 
role of the QCD contributions from the high momentum region, say above
1 GeV, we have made a check by doing the calculations with the same 
$\Sigma(p^2)$ but taking $\Lambda=1$ GeV instead of $\Lambda\to\infty$. The
results are listed in TABLE II. Comparing the nonvanishing results in
TABLE II with the corresponding $\Lambda,\bar\Lambda\to\infty$ results in 
TABLE I, we see that this change of $\Lambda$ does not cause much difference 
in $L_5,~L_7,~L_8,$ and $F_0$, while it causes $L_1,~L_2,~L_3,~L_9,$ and 
$L_{10}$ to reduce by at least a factor of 2. Therefore, we see that 
$L_5,~L_7,~L_8,$ and $F_0$ are mainly contributed by the QCD dynamics in the 
low momentum region, while high momentum region contributions to 
$L_1,~L_2,~L_3,~L_9$, and  $L_{10}$ are not negligible. 

\section{Conclusions}

In this paper, we have calculated the coefficients in the Gasser-Leutwyler
Lagrangian from the underlying theory of QCD in a simple approach with the 
approximations of taking the large-$N_c$ limit, the leading order in dynamical
perturbation theory, and the improved ladder approximation based on the QCD 
formulae given in Ref.\cite{WKWX1} to illustrate the main feature of
how QCD predicts the chiral Lagrangian coefficients. In the
calculation, we use the same regularization technique, the generalized
Schwinger proper time regularization, in the calculations of the
contributions from both the anomaly part and the normal part, so that
the relation between the contributions from the two parts can be
clearly seen.

We first take the large-$N_c$ limit to evaluate the effective action in
QCD. Our first conclusion in this study is that, {\it in the large-$N_c$ 
limit, to all orders in momentum expansion, the anomaly part contributions to 
the chiral Lagrangian coefficients [cf. Eqs.(\ref{anomalyresults})] given in 
the literature \cite{Espriu} from the effective action 
$S^{({\rm anom})}_{\rm eff}$ [cf. Eq.(\ref{Sanom})] are 
exactly cancelled by the contributions from the piece of the effective 
action $S^{({\rm norm},\Pi_{\Omega c}=0)}_{\rm eff}$ [cf. Eq.(\ref{SPi=0})] in 
the normal part contributions}, so that {\it the chiral Lagrangian 
coefficients are eventually contributed by the remaining piece of the normal 
part effective action $S^{({\rm norm},\Pi_{\Omega c}\ne 0)}_{\rm eff}$ 
[cf. Eq.(\ref{SPi})]}. The final QCD expressions for the $O(p^4)$ coefficients 
are given in Eqs.(\ref{p4full}).

To simplify  $S^{({\rm norm},\Pi_{\Omega c}\ne 0)}_{\rm eff}$, we
further make the approximation of taking the leading order in dynamical
perturbation theory. Then  $S^{({\rm norm},\Pi_{\Omega c}\ne 0)}_{\rm eff}$
is reduced to the simple form in Eq.(\ref{SGL}), and all the chiral
Lagrangian coefficients are approximately expressed in terms of the
quark self-energy $\Sigma(p^2)$ shown in Eqs.(\ref{F0B0})$\--$(\ref{Kresult}).
To solve the Schwinger-Dyson equation for $\Sigma(p^2)$, we further
take the improved ladder approximation. Lacking of the knowledge about the 
running coupling constant $\alpha_s(p)$ in the nonperturbative region, we take 
certain models for it from the literature \cite{Aoki,Munczek} [cf. 
Eqs.(\ref{omega1}), (\ref{omega2}) and (\ref{omega3})], and we further take 
the usual approximation $\alpha_s(p-q)\approx \theta(p^2-q^2)\alpha_s(p^2)
+\theta(q^2-p^2)\alpha_s(q^2)$ to simplify the calculation. The quark 
self-energy reflecting chiral symmetry breaking is obtained by solving the
simplified Schwinger-Dyson equation numerically. The obtained results of the 
$O(p^4)$ coefficients are listed in TABLE I. Compared with the experimental
values of $L_1\cdots,L_{10}$ \cite{GS}, {\it the agreement of
$L_1,~L_2,~L_4,~L_6$, and $L_{10}$ is of the level of $1\sigma$, and that
of $L_3,~L_5,~L_7$ and $L_8$ is of the level of $2\sigma$}. Only $L_9$
deviates from the experimental value by $(3\--4)\sigma$. Considering the large 
theoretical uncertainty in this simple approach, {\it all the obtained 
coefficients $L_1\cdots,L_{10}$ are consistent with the experiments}. We have 
also calculated the renormalized quark condensate 
$\langle\bar\psi\psi\rangle_r$ from the obtained $O(p^2)$ coefficient [cf. 
Eq.(\ref{condensaterenorm})] {\it which is also consistent with the experiment}.

Although the approximations in this simple approach are rather crude,
the above results show that {\it this simple approach does reveal the main
feature of QCD predictions for the chiral Lagrangian coefficients}. For
studying physics not requiring high precision, this simple approach may
already be useful. Of course further improvements of the approximations
beyond this simple approach (reflecting more about QCD dynamics) are needed. 
This kind of study is in progress and will be presented in another paper.

The approach can also be applied to electroweak theories to
study how the coefficients in the electroweak chiral Lagrangian are
predicted by various kinds of underlying gauge theories of the electroweak
symmetry breaking mechanism. This kind of study is also in progress,
and will be presented in separate papers.

\section*{Acknowledgments}

This work is supported by the National Natural science Foundation of China, 
the Foundation of Fundamental Research grant of  Tsinghua University,
and a special grant from the Ministry of Education of China.  


\appendix
\section{Functional determinant containing quark self-energy}\label{HT}

In this appendix, we take the Schwinger proper time regulation to regularize 
the one-loop functional determinant in which the quark self-energy 
$\Sigma(\bar\nabla^2)$ reflecting chiral symmetry breaking takes place.

For convenience, the evaluation is done in the Euclidean space-time,
and will be analytically continued to the Minkowskian space-time after
the evaluation. The functional determinant is complex. The imaginary part is 
just the Wess-Zumino-Witten term, and its expression in terms of $\Sigma$ has
already been given in Ref.\cite{GCM} which exactly coincides Witten's result
\cite{WZW}. The phenomenology of the Wess-Zumino-Witten term is
well-known and is not related to the main purpose of this paper. So we
shall ignore the imaginary part here, and concentrate on the evaluation
of the following real part of the functional determinant
\begin{eqnarray}                  
{\rm Re}\ln{\rm Det}[D+\Sigma(-\overline{\nabla}^2)]
&=&\frac{1}{2}{\rm Tr}\ln\bigg[[D^{\dagger}+\Sigma(-\overline{\nabla}^2)]
[D+\Sigma(-\overline{\nabla}^2)]\bigg]\nonumber\\
&=&-\frac{1}{2}\lim_{\Lambda\rightarrow\infty}
\int_{\frac{1}{\Lambda^2}}^{\infty}\frac{d\tau}{\tau}~
{\rm Tr}~e^{-\tau[\overline{E}
-\nabla^2+\Sigma^2(-\overline{\nabla}^2)
+\hat{I}_\Omega\Sigma(-\bar{\nabla}^2)+\Sigma(-\bar{\nabla}^2)\tilde{I}_\Omega
-d\!\!\! /\;\Sigma(-\overline{\nabla}^2)]}
\label{lnSigma}
\end{eqnarray}
where
\begin{eqnarray}                    
&&D\equiv\nabla\!\!\!\! /\;-s_\Omega+ip_\Omega\gamma_5,\hspace{1cm}
\nabla_{\mu}\;\equiv \partial_{\mu}-iv_{\Omega\mu}-ia_{\Omega\mu}\gamma_5
=-\nabla_{\mu}^{\dagger},
\hspace{1cm}\overline{\nabla}^{\mu}\equiv\partial^{\mu}-iv_\Omega^{\mu}(x),
\nonumber\\
\label{nabladef}
&&\overline{E}-\nabla^2
+\Sigma^2(-\overline{\nabla}^2)
+\hat{I}_\Omega\Sigma(-\bar{\nabla}^2)+\Sigma(-\bar{\nabla}^2)\tilde{I}_\Omega
-d\!\!\! /\;\Sigma(-\overline{\nabla}^2)
=[D^{\dagger}+\Sigma(-\overline{\nabla}^2)][D+\Sigma(-\overline{\nabla}^2)],
\nonumber\\
&&\hat{I}_\Omega=-ia_\Omega\!\!\!\!\!\!\!/\;\;\gamma_5-s_\Omega-ip_\Omega\gamma_5,
\hspace{2cm} 
\tilde{I}_\Omega=
-ia_\Omega\!\!\!\!\!\!\!/\;\;\gamma_5-s_\Omega+ip_\Omega\gamma_5,
\nonumber\\
&&[d\!\!\! /\;\Sigma(-\overline{\nabla}^2)]\equiv
\gamma^{\mu}[d_{\mu}\Sigma(-\overline{\nabla}^2)]=
\gamma^{\mu}\bigg(\partial_{\mu}\Sigma(-\overline{\nabla}^2)
-i[v_{\Omega\mu},\Sigma(-\overline{\nabla}^2)]\bigg),\nonumber\\
\end{eqnarray}

The matrix element in Eq.(\ref{lnSigma}) can be evaluated in the momentum 
representation
\begin{eqnarray}                          
&&\langle x|e^{-\tau[\overline{E}
-\nabla^2+\Sigma^2(-\overline{\nabla}^2)
+\hat{I}_\Omega\Sigma(-\overline{\nabla}^2)+\Sigma(-\overline{\nabla}^2)
\tilde{I}_\Omega
-d\!\!\! /\;\Sigma(-\overline{\nabla}^2)]}|x\rangle\nonumber\\
&&=\int\frac{d^4p}{(2\pi)^4}\exp\bigg\{-\tau\bigg[\overline{E}(x)
-\nabla_x^2-2ip\cdot\nabla_x+p^2
+\Sigma^2(-\overline{\nabla}^2-2ip\cdot\overline{\nabla}_x+p^2)
\nonumber\\
&&\hspace{0.5cm}
+\hat{I}_\Omega\Sigma(-\overline{\nabla}_x^2-2ip\cdot\overline{\nabla}_x+p^2)
+\Sigma(-\overline{\nabla}_x^2-2ip\cdot\overline{\nabla}_x+p^2)\tilde{I}_\Omega
-d\!\!\! /\;\Sigma(-\overline{\nabla}_x^2-2ip\cdot\overline{\nabla}_x+p^2)
\bigg]\bigg\}.
\label{momexp}
\end{eqnarray}
Then after lengthy but elementary calculations and expanding in powers
of the external sources, 
we can identify the expressions for $F^2_0,~F^2_0B_0,~{\cal K}_1^{(\rm norm)},
\cdots,~{\cal K}_{15}^{(\rm norm)}$ by comparing with the form of Eqs.(\ref{p4}),
and the obtained results in the Minkowskian space-time are just
those given in Eqs.(\ref{F0}), (\ref{F0B0}) and (\ref{Kresult}) in the text. The 
details are given in Ref.\cite{det}.

For the evaluation of the effective action $S^{(\rm anom)}_{\rm eff}$
in Eq.(\ref{Sanom}) in the Minkowskian space-time, we note that there is no 
$\Sigma(-\overline\nabla^2)$ term
in Eq.(\ref{lnSigma}), but we still have to replace $\Sigma(-\overline\nabla^2)$
by an infrared cutoff parameter $\kappa$ in Eq.(\ref{lnSigma}) to
regularize the infrared divergence. Then the momentum integration can
be explicitly carried out, and we obtain the results in Eqs.(\ref{anomK})
in the text.

\section{$\Omega$-independence of the last term in eq.(19)}

Here we show that the last term in Eq.(\ref{SPi=0}),
\begin{eqnarray}                
S_{\rm eff}^{(\bar{G})}\equiv N_c\sum^{\infty}_{n=2}{\int}d^{4}x_1\cdots 
d^4x_{n}'\frac{(-i)^{n}(N_c g_s^2)^{n-1}}{n!}
\bar{G}^{\sigma_1\cdots\sigma_n}_{\rho_1
\cdots\rho_n}(x_1,x'_1,\cdots,x_n,x'_n)
\Phi^{\sigma_1\rho_1}_{\Omega c}(x_1 ,x'_1)\cdots 
\Phi^{\sigma_n\rho_n}_{\Omega c}(x_n ,x'_n)\bigg|_{\Pi_{\Omega c}=0},
\label{3rdterm}
\end{eqnarray}
with the saddle point equation [cf. Eq.(\ref{Pieq})]
\begin{eqnarray}                    
&&\Phi^{\sigma\rho}_{\Omega c}(x,y)\bigg|_{\Pi_{\Omega c}=0}
=-i[(i\partial\!\!\! /\;+J_{\Omega})^{-1}]^{\rho\sigma}(y,x)
\label{sta1}
\end{eqnarray}
is $\Omega$-independent. The $\Omega$-rotated quantities in Eqs.(\ref{3rdterm}) 
and (\ref{sta1}) are defined by \cite{WKWX1}
\begin{eqnarray}                       
&&J_{\Omega}(x)
=[\Omega(x)P_R+\Omega^{\dagger}(x)P_L]
~[J(x)+i\partial\!\!\!\! /\;][\Omega(x)P_R+\Omega^{\dagger}(x)P_L],\nonumber\\
&&\Phi_{\Omega}^T(x,y)=[\Omega^{\dagger}(x)P_R+\Omega(x)P_L]~\Phi^T(x,y)
[\Omega^{\dagger}(y)P_R+\Omega(y)P_L],
\label{rotation}
\end{eqnarray}
and the $2n$-point Green's function
$\bar{G}^{\sigma_1\cdots\sigma_n}_{\rho_1 \cdots\rho_n}(x_1,x'_1,\cdots,
x_n,x'_n)$ is define by \cite{WKWX1}
\begin{eqnarray}                       
 &&G_{\mu_1\cdots\mu_n}^{i_1\cdots{i_n}}
 (x_1,\cdots,x_n)[\overline{\psi}^{a_1}_{{\alpha}_1}(x_1)
 (\frac{\lambda_{i_1}}{2})_{\alpha_1\beta_1}\gamma^{\mu_1}
 {\psi}^{a_1}_{{\beta}_1}(x_1)]\cdots[
\overline{\psi}^{a_n}_{{\alpha}_n}(x_n)
(\frac{\lambda_{i_n}}{2})_{\alpha_n\beta_n}
\gamma^{\mu_n}{\psi}^{a_n}_{\beta_n}(x_n)]\nonumber 
\\
&&={\int}d^{4}x'_1\cdots{d^{4}x'_n}
g_s^{n-2}\overline{G}^{\sigma_1\cdots\sigma_n}_{\rho_1
\cdots\rho_n}(x_1,x'_1,\cdots,x_n,x'_n)
\overline{\psi}^{\sigma_1}_{\alpha_1}(x_1){\psi}^{\rho_1}
_{\alpha_1}(x'_1)\cdots\overline{\psi}^{\sigma_n}
 _{\alpha_n}(x_n){\psi}^{\rho_n}_{\alpha_n}(x'_n).
\label{barGdef}
\end{eqnarray}

First we see from Eqs.(\ref{rotation}) that Eq.(\ref{sta1}) can be written as
\begin{eqnarray}                    
&&\Phi^{\sigma\rho}_{\Omega c}(x,y)\bigg|_{\Pi_{\Omega c}=0}
=(\Phi^T)^{\rho\sigma}_{\Omega c}(y,x)\bigg|_{\Pi_{\Omega c}=0}
\nonumber\\
&&=\bigg[[\Omega^{\dagger}(y)P_R+\Omega(y)P_L]~\Phi^T_c(y,x)
[\Omega^{\dagger}(x)P_R+\Omega(x)P_L]\bigg]
^{\rho\sigma}\bigg|_{\Pi_c=0}\nonumber\\
&&=-i\bigg[[\Omega^{\dagger}(y)P_R+\Omega(y)P_L]~
[(i\partial\!\!\! /\;+J)^{-1}](y,x)
[\Omega^{\dagger}(x)P_R+\Omega(x)P_L]\bigg]^{\rho\sigma}\nonumber\\
&&=-i[\Omega^{\dagger}(y)P_R+\Omega(y)P_L]^{\rho\rho'}
[(i\partial\!\!\! /\;+J)^{-1}]^{\rho'\sigma'}(y,x)
[\Omega^{\dagger}(x)P_R+\Omega(x)P_L]^{\sigma'\sigma}\nonumber\\
&&=-i[\gamma_0V_{\Omega}\gamma_0]^{\dagger\rho\rho'}(y)
[(i\partial\!\!\! /\;+J)^{-1}]^{\rho'\sigma'}(y,x)
V_{\Omega}^{\sigma'\sigma}(x),
\label{PhiO}
\end{eqnarray}
in which
\begin{eqnarray}                 
V_{\Omega}(x)\equiv\Omega^{\dagger}(x)P_R+\Omega(x)P_L
\label{Vdef}
\end{eqnarray}
satisfies
\begin{eqnarray}                         
\gamma_0V_\Omega^{\dagger}(x)\gamma_0\gamma_{\mu}=\gamma_{\mu}
V_\Omega^{\dagger}(x),
\label{Vcondition}
\end{eqnarray}
and 
\begin{eqnarray}                           
\Phi^T_c(y,x)\bigg|_{\Pi_c=0}=-i[(i\partial\!\!\! /\;+J)^{-1}](y,x)
\label{Phic}
\end{eqnarray}
is $\Omega$-independent.

With the expression (\ref{PhiO}) for 
$\Phi^{\sigma\rho}_{\Omega c}(x,y)|_{\Pi_{\Omega c}=0}$, 
Eq.(\ref{3rdterm}) becomes
\begin{eqnarray}                 
S^{(\bar{G})}_{\rm eff}&&=N_c\sum^{\infty}_{n=2}{\int}d^{4}x_1\cdots d^4x_{n}'
\frac{(-i)^{n}(N_c g_s^2)^{n-1}}{n!}\bar{G}^{\sigma_1\cdots\sigma_n}_{\rho_1
\cdots\rho_n}(x_1,x'_1,\cdots,x_n,x'_n)
\Phi^{\sigma_1\rho_1}_{\Omega c}(x_1 ,x'_1)\cdots 
\Phi^{\sigma_n\rho_n}_{\Omega c}(x_n ,x'_n)\bigg|_{\Pi_{\Omega c}=0}\nonumber\\
&&=N_c\sum^{\infty}_{n=2}{\int}d^{4}x_1\cdots d^4x_{n}'
\frac{(-i)^{n}(N_c g_s^2)^{n-1}}{n!}\bar{G}^{\sigma_1\cdots\sigma_n}_{\rho_1
\cdots\rho_n}(x_1,x'_1,\cdots,x_n,x'_n)
(\gamma_0V_{\Omega}^{\dagger}\gamma_0)^{\sigma_1\sigma'_1}(x_1)\cdots
\nonumber\\
&&\hspace{0.4cm}\times(\gamma_0V_{\Omega}^{\dagger}\gamma_0)^{\sigma_n
\sigma'_n}(x_n)V_{\Omega}^{\rho'_1\rho_1}(x'_1)\cdots 
V_{\Omega}^{\rho'_n\rho_n}(x'_n)\Phi^{\sigma_1\rho_1}_c(x_1 ,x'_1)\cdots 
\Phi^{\sigma_n\rho_n}_c(x_n ,x'_n)\bigg|_{\Pi_c=0}\nonumber\\
&&=N_c\sum^{\infty}_{n=2}{\int}d^{4}x_1\cdots d^4x_{n}'
\frac{(-i)^{n}(N_c g_s^2)^{n-1}}{n!}\bar{G}
^{\sigma_1'\cdots\sigma_n'}_{V_\Omega,
\rho'_1\cdots\rho'_n}(x_1,x'_1,\cdots,x_n,x'_n)
\Phi^{\sigma_1'\rho_1'}_c(x_1 ,x'_1)\cdots 
\Phi^{\sigma_n'\rho_n'}_c(x_n ,x'_n)\bigg|_{\Pi_c=0},
\label{SGV}
\end{eqnarray}
where $\bar{G}_{V_\Omega,\rho'_1\cdots\rho'_n}^{\sigma'_1\cdots\sigma'_n}
(x_1,x'_1,\cdots,x_n,x'_n)$ is 
\begin{eqnarray}                     
\bar{G}^{\sigma_1\cdots\sigma_n}_{V_\Omega,\rho_1\cdots\rho_n}
(x_1,x'_1,\cdots,x_n,x'_n)
&\equiv& (\gamma_0V^{\dagger}_\Omega\gamma_0)^{\sigma_1\sigma'_1}(x_1)\cdots
(\gamma_0V^{\dagger}_\Omega\gamma_0)^{\sigma_n\sigma'_n}(x_n)\nonumber\\
&&\times\bar{G}^{\sigma'_1\cdots\sigma'_n}_{\rho'_1\cdots\rho'_n}
(x_1,x'_1,\cdots,x_n,x'_n)V_\Omega^{\rho'_1\rho_1}(x'_1)\cdots V_\Omega^{\rho'_n\rho_n}
(x'_n).
\label{GV}
\end{eqnarray}

Next, we look at this transformed Green's function $\bar{G}^{\sigma_1\cdots
\sigma_n}_{V_\Omega,\rho_1\cdots\rho_n}
(x_1,x'_1,\cdots,x_n,x'_n)$. From the definition (\ref{barGdef}) and the 
property (\ref{Vcondition}) we have
\begin{eqnarray}                      
&&{\int}d^{4}x'_1\cdots{d^{4}x'_n}
g_s^{n-2}\bar{G}^{\sigma_1\cdots\sigma_n}_{V_\Omega,\rho_1
\cdots\rho_n}(x_1,x'_1,\cdots,x_n,x'_n)
\overline{\psi}^{\sigma_1}_{\alpha_1}(x_1){\psi}^{\rho_1}
_{\alpha_1}(x'_1)\cdots\overline{\psi}^{\sigma_n}
 _{\alpha_n}(x_n){\psi}^{\rho_n}_{\alpha_n}(x'_n)\nonumber\\
&=&{\int}d^{4}x'_1\cdots{d^{4}x'_n}
g_s^{n-2}\bar{G}^{\sigma_1\cdots\sigma_n}_{\rho_1
\cdots\rho_n}(x_1,x'_1,\cdots,x_n,x'_n)
(\overline{\psi}\gamma_0V^{\dagger}_\Omega\gamma_0)^{\sigma_1}_{\alpha_1}(x_1)
(V_\Omega{\psi})^{\rho_1}_{\alpha_1}(x'_1)\cdots\nonumber\\
 && \cdots(\overline{\psi}\gamma_0V^{\dagger}_\Omega\gamma_0)^{\sigma_n}
 _{\alpha_n}(x_n)(V_\Omega{\psi})^{\rho_n}_{\alpha_n}(x'_n)\nonumber\\
 &=&G_{\mu_1\cdots\mu_n}^{i_1\cdots{i_n}}
 (x_1,\cdots,x_n)[(\overline{\psi}\gamma_0V^{\dagger}_\Omega
 \gamma_0)^{a_1}_{{\alpha}_1}(x_1)
 (\frac{\lambda_{i_1}}{2})_{\alpha_1\beta_1}\gamma^{\mu_1}
 (V_\Omega{\psi})^{a_1}_{{\beta}_1}(x_1)]\cdots\nonumber\\
&&\cdots[(\overline{\psi}\gamma_0V^{\dagger}_\Omega\gamma_0)^{a_n}_{{\alpha}_n}(x_n)
(\frac{\lambda_{i_n}}{2})_{\alpha_n\beta_n}
\gamma^{\mu_n}(V_\Omega{\psi})^{a_n}_{\beta_n}(x_n)]\nonumber 
\\
 &=&G_{\mu_1\cdots\mu_n}^{i_1\cdots{i_n}}
 (x_1,\cdots,x_n)[\overline{\psi}^{a_1}_{{\alpha}_1}(x_1)
 (\frac{\lambda_{i_1}}{2})_{\alpha_1\beta_1}\gamma^{\mu_1}
 {\psi}^{a_1}_{{\beta}_1}(x_1)]\cdots[
\overline{\psi}^{a_n}_{{\alpha}_n}(x_n)
(\frac{\lambda_{i_n}}{2})_{\alpha_n\beta_n}
\gamma^{\mu_n}{\psi}^{a_n}_{\beta_n}(x_n)]\nonumber 
\\
&=&{\int}d^{4}x'_1\cdots{d^{4}x'_n}
g_s^{n-2}\bar{G}^{\sigma_1\cdots\sigma_n}_{\rho_1
\cdots\rho_n}(x_1,x'_1,\cdots,x_n,x'_n)
\overline{\psi}^{\sigma_1}_{\alpha_1}(x_1){\psi}^{\rho_1}
_{\alpha_1}(x'_1)\cdots\overline{\psi}^{\sigma_n}
 _{\alpha_n}(x_n){\psi}^{\rho_n}_{\alpha_n}(x'_n),\nonumber
\end{eqnarray}                                   
i.e.,
\begin{eqnarray}                   
\bar{G}^{\sigma_1\cdots\sigma_n}_{V_\Omega,\rho_1
\cdots\rho_n}(x_1,x'_1,\cdots,x_n,x'_n)
=\overline{G}^{\sigma_1\cdots\sigma_n}_{\rho_1
\cdots\rho_n}(x_1,x'_1,\cdots,x_n,x'_n). 
\label{GV}
\end{eqnarray}
Thus the transformed Green's function 
$\bar{G}_{V_\Omega\rho_1\cdots\rho_n}^{\sigma_1\cdots\sigma_n}$
in Eq.(\ref{SGV}) can be replaced by $\bar{G}_{\rho_1\cdots\rho_n}^{\sigma_1\cdots
\sigma_n}$, and Eq.(\ref{SGV}) becomes
\begin{eqnarray}                
S^{(\bar{G})}_{\rm eff}&=&N_c\sum^{\infty}_{n=2}{\int}d^{4}x_1\cdots d^4x_{n}'
\frac{(-i)^{n}(N_c g_s^2)^{n-1}}{n!}\bar{G}
^{\sigma_1\cdots\sigma_n}_{\rho_1\cdots\rho_n}(x_1,x'_1,\cdots,x_n,x'_n)
\Phi^{\sigma_1\rho_1}_c(x_1 ,x'_1)\cdots 
\Phi^{\sigma_n\rho_n}_c(x_n ,x'_n)\bigg|_{\Pi_c=0}
\end{eqnarray}
which is independent of $\Omega$.


\newpage
\begin{center}
{\large\bf Tables}\end{center}
\begin{table}[t]
\null\noindent
{\small{\bf TABLE I}. The obtained values of the $O(p^4)$ coefficients 
$L_1\cdots,L_{10}$ for Model A [Eq.(\ref{omega1})], Model B 
[Eq.(\ref{omega2})] and Model C [Eq.(\ref{omega3})] with
$\Lambda,\bar\Lambda\to\infty$ together with the
experimental values [Eq.(\ref{exptL})] for comparison. $\Lambda_{QCD}$
is in MeV, and the coefficients are in units of $10^{-3}$.}
\begin{tabular}{c c c c c c c c c c c c}
 &$\Lambda_{QCD}$&$L_1$ &$L_2$ &$L_3$ &$L_4$ & $L_5$ & $L_6$ & $L_7$ &
$L_8$ & $L_9$ & $L_{10}$\\
\hline
A: &484& 1.10 &2.20 &-7.82 &0 &1.62 & 0 & -0.70&
1.75&5.07&-7.06\\ 
B: &230 & 0.921 & 1.84 & -6.73 & 0 & 1.43 & 0 & -0.673 & 1.64 & 3.80 & 
-6.22\\ 
C: &230 & 0.948 & 1.90 & -6.90 & 0 & 1.29 & 0 & -0.632 & 1.56 & 
3.95 & -6.21 \\
Expt:& &$0.9\pm 0.3$&$1.7\pm 0.7$&$-4.4\pm 2.5$&$0\pm 0.5$&$2.2\pm
0.5$&$0\pm 0.3$&$-0.4\pm 0.15$&$1.1\pm 0.3$&$7.4\pm 0.7$&$-6.0\pm 0.7$\\
\end{tabular}
\end{table}

\vspace{1cm}
\begin{table}[h] 
\null\noindent
{\small{\bf TABLE II}. The same as in TABLE I but with $\Lambda=1$ GeV 
instead of $\Lambda\to\infty$.}
\begin{tabular}{c c c c c c c c c c c c }
 &$L_1$ &$L_2$ &$L_3$ &$L_4$ & $L_5$ & $L_6$ & $L_7$ &
$L_8$ & $L_9$ & $L_{10}$ & $F_0$ \\
\hline
A: &0.403&0.805&-3.47&0 &1.47& 0
&-0.792&1.83&2.28&-4.08&88.7 \\ 
B: &0.281&0.563&-2.71&0 &1.44& 0 &-0.836&1.83&1.46&-3.69&89.6 \\ 
C: &0.304&0.608&-2.86&0 &1.43& 0 &-0.855&1.87&1.56&-3.64&89.4 \\
\end{tabular}
\end{table}

\begin{center}
{\large\bf Figures}
\end{center}
\vspace{-2cm}
\centerline{\epsffile{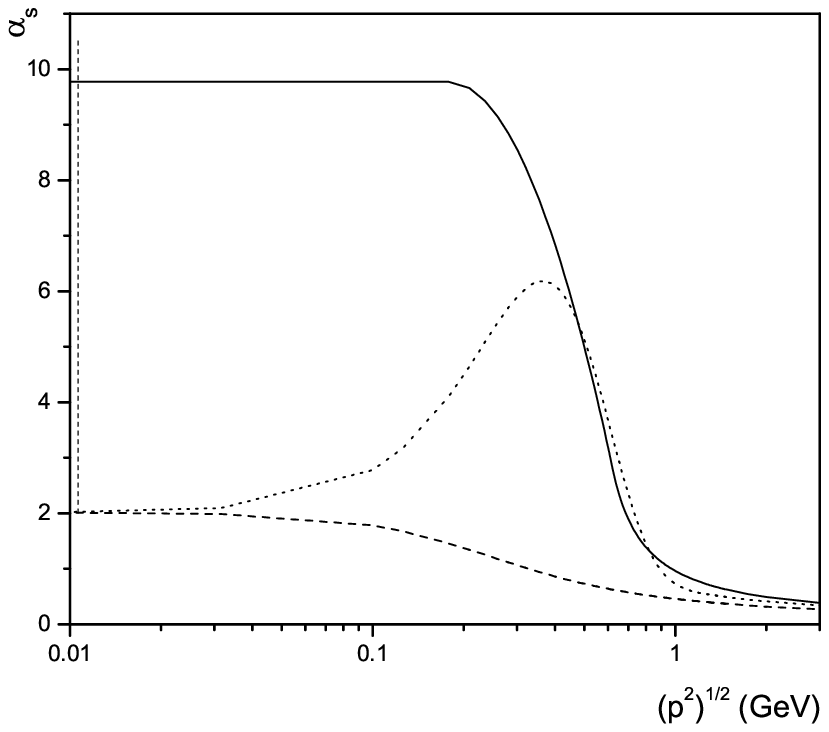}} 
\noindent FIG.1. $\alpha_s(p)$ for Model A [Eq.(\ref{omega1})], Model B
[Eq.(\ref{omega2})], and Model C [Eq. (\ref{omega3})]. 
The solid, dashed, and dotted lines are for Models A, B, and C, respectively.

\vspace{0.4cm}
\centerline{\epsffile{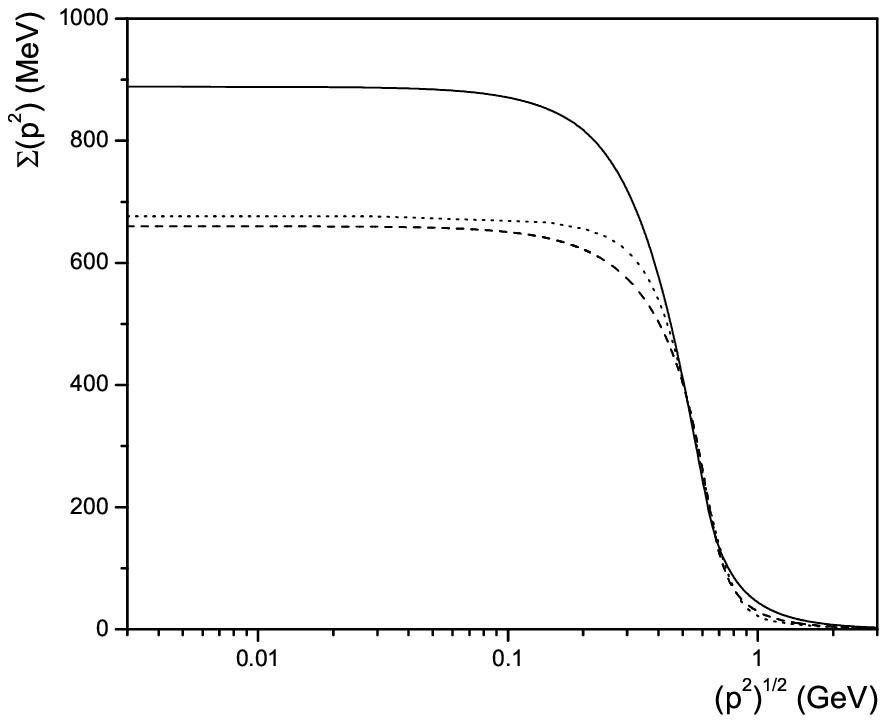}} 
\noindent FIG. 2. The obtained $\Sigma(p^2)$ from the Schwinger-Dyson
equation (\ref{eq0}) with Model A [Eq.(\ref{omega1})],
Model B [Eq.(\ref{omega2})] and Model C [Eq. (\ref{omega3})] for the running 
coupling constant $\alpha_s$. The solid, dashed, and dotted lines are for 
Models A, B, and C, respectively.

\end{document}